\documentclass[prb,twocolumn,showpacs,preprintnumbers,amsmath,amssymb, superscriptaddress]{revtex4-1}

\usepackage{lmodern}
\usepackage[makeroom]{cancel}
\usepackage{amsmath}    
\usepackage{amssymb}
\usepackage{graphicx}   
\usepackage{verbatim}   
\usepackage{color}      
\usepackage{hyperref}   
\usepackage[normalem]{ulem}
\usepackage{natbib}
\usepackage{fixmath}
\usepackage{enumitem}
\usepackage{ulem}
\usepackage{youngtab}

\usepackage{soul}   

\def \ket #1{\left| #1 \right\rangle}
\def \bra #1{\left\langle #1 \right|}

\def \scalarprod #1#2{\left\langle #1 | #2 \right\rangle}
\def \cgcoeff #1#2{\left( #1 | #2 \right)}
\def \loc {\mathrm{loc}}
\def \brc #1{\left\lbrace #1 \right\rbrace}

\hypersetup{colorlinks,linkcolor=blue,urlcolor=blue,citecolor=blue}
\newcommand{\beq}{\begin{equation}}
\newcommand{\eeq}{\end{equation}}
\newcommand{\bea}{\begin{eqnarray}}
\newcommand{\eea}{\end{eqnarray}}

\newcommand{\be}{\begin{equation}}
\newcommand{\ee}{\end{equation}}

\definecolor{darkgreen}{rgb}{0,0.5,0}
\definecolor{orange}{rgb}{1,0.5,0}
\definecolor{grey}{rgb}{.6,.6,.6}

\begin{document}
\title{Quantum Quench and Charge Oscillations  in the SU(3) Hubbard Model: a Test of Time Evolving Block Decimation with 
general non-Abelian Symmetries}
        
\author{Mikl\'os Antal Werner}
\affiliation{BME-MTA Exotic Quantum Phases 'Lend\"ulet' Research Group, Institute of Physics, Budapest University of Technology and Economics, 
Budafoki \'ut 8., H-1111 Budapest, Hungary}
\affiliation{MTA-BME Quantum Dynamics and Correlations Research Group, 
Institute of Physics, Budapest University of Technology and Economics, 
Budafoki \'ut 8., H-1111 Budapest, Hungary}
\author{C\u at\u alin Pa\c scu Moca}
\affiliation{MTA-BME Quantum Dynamics and Correlations Research Group, 
Institute of Physics, Budapest University of Technology and Economics, 
Budafoki \'ut 8., H-1111 Budapest, Hungary}
\affiliation{Department of Physics, University of Oradea, 410087, Oradea, Romania}

\author{\" Ors Legeza}
\affiliation{Strongly Correlated Systems 'Lend\" ulet' Research Group, Institute for Solid State Physics and Optics,
 MTA Wigner Research Centre for Physics, P.O. Box 49, H-1525 Budapest, Hungary}	
\author{Gergely Zar\'and}
\affiliation{BME-MTA Exotic Quantum Phases 'Lend\"ulet' Research Group, Institute of Physics, Budapest University of Technology and Economics, 
Budafoki \'ut 8., H-1111 Budapest, Hungary}
\affiliation{MTA-BME Quantum Dynamics and Correlations Research Group, 
Institute of Physics, Budapest University of Technology and Economics, 
Budafoki \'ut 8., H-1111 Budapest, Hungary}
\date{\today}
\begin{abstract}
We introduce the notion of non-Abelian tensors, and  use them to construct a general non-Abelian 
time evolving block decimation
(NA-TEBD) scheme that uses an arbitrary number of Abelian and non-Abelian symmetries.
Our approach increases the speed  and memory storage efficiency of matrix product state based
 computations by several orders of magnitudes, and makes large bond dimensions accessible even
 on simple desktop architectures. We use it to study post-quench dynamics in the repulsive SU(3) Hubbard model, and to 
determine the time evolution of various local operators and correlation functions efficiently. 
Interactions turn algebraic charge relaxation into exponential, and  suppress 
coherent quantum oscillations rapidly.
\end{abstract}
\maketitle

\section{Introduction}\label{sec:Introduction}

Matrix product state based numerical renormalization approaches such as Wilson's original numerical 
renormalization group (NRG) method~\cite {Wilson_NRG, Wilson_NRG2} or the density matrix renormalization 
group (DMRG) introduced by Steven R. White,\cite {White_DMRG, Schollwoeck_Rev2005, Schollwoeck_Rev2010} 
 proved to be extremely powerful tools to study low-energy properties of strongly interacting many-body systems. 

Though the original methods were designed to address the ground state properties of zero and one dimensional 
quantum systems, modern offsprings of  NRG and DMRG  earned wide applications: the time-evolving block decimation (TEBD) algorithm,\cite {Vidal2004, Vidal2007}  time dependent DMRG,\cite{White_tDMRG_2004}
 or the time dependent variational principle (TDVP) algorithms~\cite {Haegeman}
allow one to study  the   evolution of closed quantum systems  in real or imaginary time, while  
in two dimensions,   the projected entangled paired states (PEPS) approach~\cite{Verstraete_PEPS, Orus} has been proposed 
as a viable extension of MPS states.  For Gapless models Multiscale Entanglement Renormalization Ansatz (MERA) is a suitable choice,\cite{Vidal_MERA} while tree tensor network  states  (TTNS) represent another promising 
direction for models with long-ranged interactions.~\cite {Klaas,Chan}

DMRG, however,  continues to be a very attractive and robust approach for systems with long-ranged interactions 
as well as for one- and two-dimensional systems,  (see, e.g., Ref.~\onlinecite {Szalay}), and provides  a valid alternative to 
more sophisticated approaches, which often display  less favorable 
computational scaling with the so-called bond dimension.\cite{Lubasch_2014, Corboz_2016}

Exploiting the symmetry of the problem as much as one can is always a crucial   ingredient 
in numerical simulations: it  reduces the computational   cost  and  boosts up the accuracy. 
It is  straightforward to implement Abelian symmetries, such as parity or charge conservation
in  most MPS and tensor network algorithms.\cite{iTENSOR, ALPS}  Handling 
non-Abelian symmetries is, however, much more challenging. It has been known for a long time 
how  to treat non-Abelian symmetries in NRG~\cite{Wilson_NRG, Wilson_NRG2, TothAnnaNRG, PascuNRG} and 
 DMRG\cite{GulacsiMcCulloch}  simulations,  and  $SU(2)$ symmetry has also been implemented in 
 TEBD,\cite{SinghVidal2010}  in TTNS,\cite{Gunst2019} and in PEPS,\cite{Hubig2018}
 yet  a unified non-Abelian tensor framework incorporating non-Abelian symmetries
{remained a challenge.}

 \begin{figure}[b!]
\begin{center}
 \includegraphics[width=0.3\columnwidth]{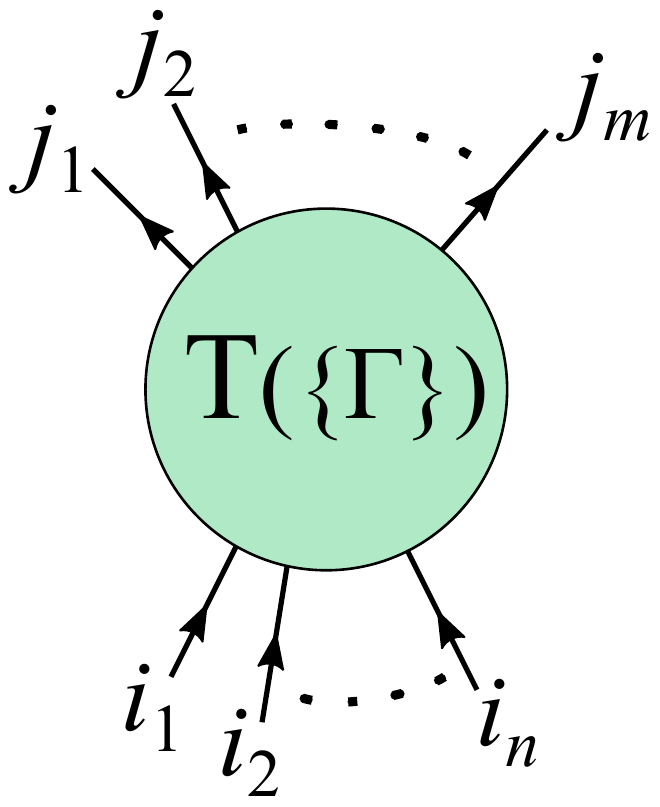}
 \caption{Graphical representation of an  NA-tensor $ T (\brc {\Gamma})_{i_1 \, i_2 \dots i_n}^{j_1 \, j_2 \dots j_m}$. Incoming and outgoing  legs correspond to  lower and upper  indices. Tensor blocks {are labeled by} the representation indices $ \brc {\Gamma} = (\Gamma_1, \dots, \Gamma_k) $.}
 \label{fig:NAtensor}
\end{center}
 \end{figure}
 
In Ref.~\onlinecite{S1_Miklos},  we   introduced  the general structure of non-Abelian tensors 
(NA-tensors), which provides the requested unified framework, and applied this approach  to describe the 
time evolution of the $S=1$ Heisenberg chain and to study quasiparticle dynamics.  NA-tensors, depicted in Fig.1, 
are objects that carry symmetry labels ({representation labels}) as internal arguments, and have external legs, which, however, may be tied to the aforementioned  internal symmetry labels. 
Line directions indicate  regular or conjugate representations. 
We  remark that our NA-tensors resemble the previously introduced Q-spaces tensor class,\cite{Weichselbaum2012, Weichselbaum2020} but  the treatment of symmetry dependent parts of the tensor networks is substantially 
and conceptually different in our construction.
The structure we introduce here 
encodes in a natural way non-Abelian MPS structures, Clebsch-Gordan coefficients, 6J and 9J symbols,\cite{Wigner} 
and provides a technically transparent framework to handle non-Abelian symmetries, in general.

In this work, we give a detailed account of this mathematical and computational 
 framework, and demonstrate its performance  on an experimentally relevant system,
 the fermionic SU(3) Hubbard model, 
\begin{equation}
\hat H = - J \sum_{\alpha}\sum_{l=1}^{L-1}\big ( c^\dagger_{l,\alpha}c_{l+1,\alpha} +h.c.\big)
+ U \sum_{l=1}^L\sum_{\alpha\ne \alpha'} 
 n_{l,\alpha}n_{l,\alpha'}\, .  
\label{eq:SU3}
\end{equation}
Here $J$ denotes the hopping amplitude between nearest-neighbors sites, $U$ represents 
the local strength of the interaction and $n_{i, \alpha}$ is the number operator at a given site,
$n_{i, \alpha} =c^\dagger_{i,\alpha}c_{i,\alpha} $. This model  displays an overall 
$\text {SU(3)}\times \text{U(1)}$ symmetry, which we use to obtain a compact NA-MPS 
description of the time evolution. In the following,
 if not explicitly displayed, energies and time are measured in units of $J$ and $J^{-1}$, respectively. 
  
\begin{figure}[t!]
	\includegraphics[width=0.85\columnwidth]{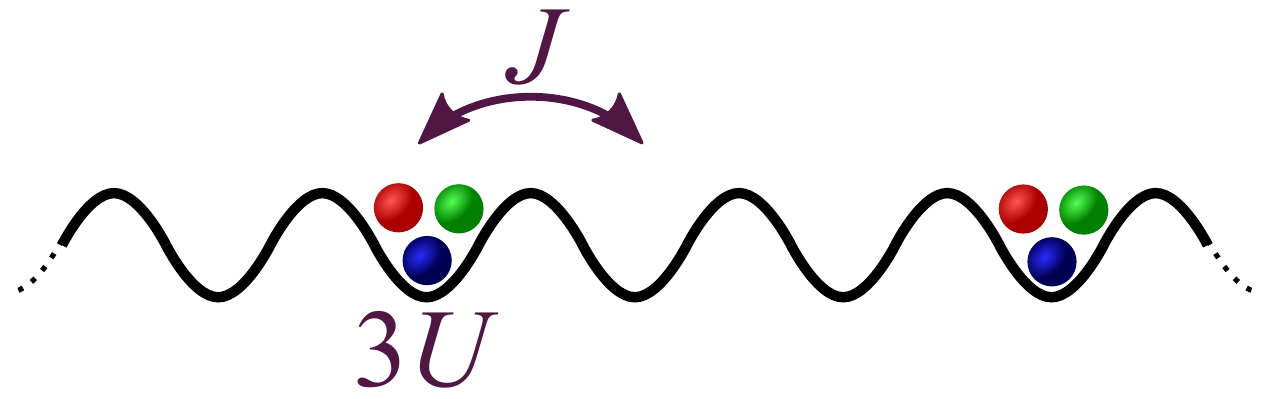}
\caption{Initial state of the SU(3) Hubbard chain. Fermions sit in groups of 3 on every third site of the 
optical lattice. 
 }
	\label{fig:quench_sketch}
\end{figure}

The one-dimensional model,  Eq.~\eqref{eq:SU3} is not just of pure theoretical interest. Both its attractive~\cite{Hofstetter} 
and  its repulsive versions~\cite{Boll_2016} have been realized by ultracold atoms trapped in optical lattices,
where the real-time  dynamics can be carefully observed. Here we study the SU(3) version of the experiment 
realized in  Ref~\onlinecite{Trotzky2012}  with $^{87}\mathrm{Rb}$ atoms:  we prepare an initial state with 
groups of three atoms   placed on every third site (see Fig.\ref{fig:quench_sketch}). 
As we demonstrate, in the absence of interactions
 charge oscillations relax  to the average occupation algebraically,  
and long-ranged spatial correlations develop. A finite interaction strength changes this behavior 
dramatically, rapidly suppresses coherent charge oscillations, 
and induces  exponential charge equilibration.  

The paper is structured as follows: {We introduce non-Abelian matrix product states in  Section \ref{sec:MPS_nonsym}. 
Non-Abelian tensors (NA-tensors) and their algebraic properties are presented in Section \ref{sec:NA-tensor}.
We describe the generalized non-Abelian TEBD algorithm
 in Section \ref{sec:NATEBD},  while results for various quantities such as the charge oscillation or the
entanglement entropy growth in the  SU(3) Hubbard model are presented in  Section \ref{sec:Hubbard}. 
We  benchmark the efficiency of our code In Section \ref{sec:Numerical_test},
and summarize our results  and conclusions in Section \ref{sec:Conclusions}.
Certain technical details  have been relegated to appendices.  
}

\section{Non-Abelian Matrix Product States }
\label{sec:MPS_nonsym}
\subsection{MPS representation of  quantum states}
\label{subsec:MPS_nonsym}
The MPS representation of a state $ \ket {\Psi} $  can be written as\cite{Schollwoeck_Rev2010,Orus} 
\begin{eqnarray}
 \ket{\Psi} &=& \sum_{a_{1}, \dots a_{L-1}} \sum_{\sigma_1, \dots \sigma_L} \mathcal{A}^{[1] \, a_1}_{\sigma_1} \mathcal{A}^{[2] \, a_2}_{a_1 \sigma_2} \dots \mathcal{A}^{[L]}_{a_{L-1} \sigma_L}  \times \nonumber \\
 & & \ket{\sigma_1} \otimes \ket{\sigma_2} \otimes \dots \otimes \ket{\sigma_L} \; .
 \label{eq:left_MPS_nonsym_full}
\end{eqnarray}
Here the states $\ket{\sigma_l}$ span the local Hilbert space $\mathcal{H}_l$ at site $l$.
In case of {the SU(3)} Hubbard chain, e.g., each site has $2^3 = 8$ states. 
{Representation Eq.~\eqref{eq:left_MPS_nonsym_full} possesses an
enormous gauge freedom, and the   'matrices'  $ \mathcal{A}^{[l]}$ are not uniquely defined. In the following, we  
use the so-called 'left-canonical' MPS representation, where the MPS is obtained by using the left Schmidt states of the 
so-called Schmidt decomposition.\cite{Schollwoeck_Rev2010, Orus,  Schmidt_reference}
To achieve this,  we cut the system into two parts at bond $l$, and  perform a Schmidt decomposition with this partitioning to yield
\begin{equation}\label{eq:Schmidt_decomp_nonsym}
\ket{\Psi} \to \sum_{a}  \lambda^{[l]}_a \, \ket{a}_l \otimes \ket{\bar a}_{ l}\;.
\end{equation}
Here $\ket{a}_l $ and $ \ket{\bar a}_l$ refer to the left and right orthonormal Schmidt states, respectively. Making a cut at the bond $l+1$ yields a similar decomposition with another set of left Schmidt states, 
$\ket{a}_{l+1}$. These latter  can, however, also be built up from  the states $\ket{a}_{l}$ and the 
local states $\ket{\sigma}_{l+1}$   at site $l+1$, as
\begin{equation}
\ket{a'}_{l+1} = \sum_{a ,\sigma} \big(\mathcal{A}^{[l+1]}\big)^{ a'}_{a \sigma} \ket{a}_{l} \otimes \ket{\sigma}_{l+1} \; .
 \label{eq:recursion}
\end{equation}
Iterating this equation leads to the MPS representation, Eq.~\eqref{eq:left_MPS_nonsym_full}. Due to the orthogonality of Schmidt states the $ \mathcal {A}$'s
satisfy the 'half-unitary' conditions, 
\begin{equation}\label{eq:MPS_nonsym_orthogonality}
 \sum_{\sigma, a} \mathcal{A}_{\phantom{ab} a\, \sigma}^{[l]\, a'} \left(\mathcal{A}_{\phantom{ab} a\, \sigma}^{[l]\, a'} \right)^{*} = \delta_{a'}^{a'}\;.
\end{equation}
One can similarly introduce 'right-canonical' {MPS} based on the right Schmidt states $\ket{\bar{a}}_l$, however, since TEBD can be formulated purely in terms 
of left-canonical {matrices}, we do not discuss them here. 
}

\begin{figure}[tb!]
	\includegraphics[width=0.65\columnwidth]{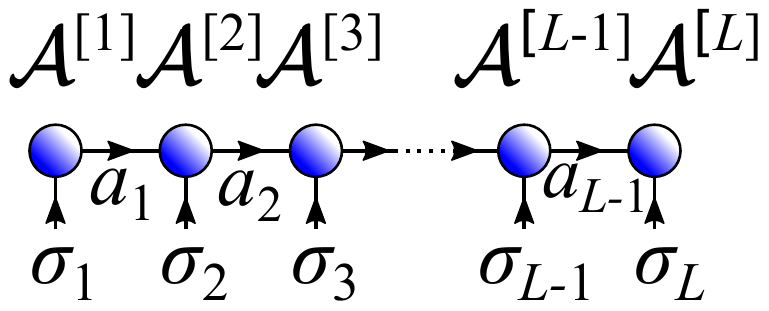}
\caption{Left-canonical MPS tensor diagram corresponding to  Eq.~\eqref{eq:left_MPS_nonsym_full}.  }	
	\label{fig:MPS_demo}
\end{figure}

Notice that the 'matrices'  $ \mathcal{A}^{[l]}$ are rather tensors than matrices, since they have three indices. 
The two 'incoming' states $\ket{a}_{l}$ and $\ket{\sigma}_{l}$ in Eq.~\ref{eq:recursion} appear
as lower indices,  while  the 'outgoing' state $\ket{a'}_{l+1}$ is displayed as an upper  index.
 This leads us to the 
 pictorial representation in Fig.~\ref{fig:MPS_demo}. It is useful to associate  incoming arrows  with
  lower ('ket') indices, and 
  outgoing arrows  with upper  ('bra') indices. The aforementioned gauge symmetry implies namely
  that incoming legs can only be contracted with 'outgoing' ones.

\subsection{Locally generated global  symmetries for  lattice models}

Generic Hamiltonians display various symmetries, which help us to organize  states. 
Here we consider  symmetries with unitary representations, 
where each element $g$ of a symmetry group $ \mathcal {G} $ 
corresponda  to some unitary operator, 
$\hat U(g)$, which commutes with the Hamiltonian, 
\begin{equation}\label{eq:global_sym_commutator}
 \left[\hat{H},\hat{\mathcal{U}}(g) \right] = 0\, , \quad \forall g \in \mathcal{G} \; .
\end{equation}
Eigenstates of the   Hamiltonian can then be labeled by the irreducible representations ('quantum numbers') 
of the  group $ \mathcal {G}$, 
 and can thus be organized into multiplets, 
\begin{equation}
 \mathcal{H} = \mathrm{span}\left\lbrace \ket{\Gamma; t_{\Gamma},m_{\Gamma}} \right\rbrace \;.
\end{equation}
Here we have grouped states into 'sectors' according to to the representation index $ \Gamma $.
Within each sector $\Gamma$,   $ t_\Gamma $ labels  the multiplets, while  $ m_\Gamma $ is a symmetry-related internal quantum number. {States within a multiplet  $(\Gamma; t_\Gamma)$  are transformed among each other under the action 
of the $\hat{\mathcal{U}}(g)$'s,  and are degenerate.}

For multiple symmetries, that is $ \mathcal {G} = \mathcal {G}_1 \otimes \dots \otimes \mathcal {G}_{n_S} $,  representation indices form a list $ \Gamma = (\Gamma_1, \dots , \Gamma_ {n_S})$.
In case of the SU(3) Hubbard model, discussed here,  the global symmetry is SU(3) $\times$U(1), and, accordingly, 
multiplets will be labeled by SU(3) representations and U(1) charges (i.e., particle number).

Here we restrict ourselves to  \emph{locally generated} global symmetries,  for which  $ \hat {\mathcal {U}} (g) $ 
factorizes as
\begin{equation}
 \hat{\mathcal{U}}(g) = \hat{\mathcal{U}}_1(g) \otimes \hat{\mathcal{U}}_2(g) \otimes \dots \otimes \hat{\mathcal{U}}_L(g)\, ,
\end{equation}
 with the $ \hat {\mathcal {U}}_i (g) $'s operating only at site $ i $. In this case , the \emph{local}  Hilbert space  at each lattice site $i$ can also be organized into multiplets (sectors), 
\begin{equation}
 \mathcal{H}_i = \mathrm{span}\left\lbrace \ket{\Gamma^\loc; \tau_{\Gamma^\loc}, \mu_{\Gamma^\loc}}_i \right\rbrace \; ,
\end{equation}
with the $ \Gamma^ {\loc} $ labeling local representations, $ \tau_{\Gamma^\loc} $ denoting the associated local 
multiplets, and  $ \mu_{\Gamma^\loc}$  the internal index of the given representation. 

Hereinafter, for clarity,   multiplet and internal labels associated with a \emph{single} site 
shall be  denoted by \emph{ greek letters}, while  states or multiplets for multi-site (sub) systems are denoted by \emph{latin letters}.

\subsection{Non-belian MPS: matrix product states with non-Abelian symmetries}\label{subsection:NA-MPS}

The easiest way to obtain the non-Abelian MPS (NA-MPS) representation of a state  $ \ket {\Psi} $
is to exploit Schmidt decomposition introduced in Subsection~\ref{subsec:MPS_nonsym}. The construction 
 in Eqs.~\eqref{eq:Schmidt_decomp_nonsym} and \eqref{eq:recursion} carries over in the 
 presence of non-Abelian symmetries, too. The only modification is that 
Schmidt states are now grouped into multiplets, $\ket{a}\to \ket{\Gamma; t_\Gamma,m_\Gamma} $, 
and  Schmidt states constructed on neighboring bonds $l$ and $l+1$ are related via 
the Clebsch-Gordan coefficients of the symmetry group $\mathcal{G}$, 
\begin{widetext}
\begin{eqnarray} \label{eq:NAMPS_recursion}
  \ket{\Gamma'; t_{\Gamma'}, m_{\Gamma'}}_{l+1} =& & \sum_{\Gamma,\Gamma^\loc} \; \sum_{t_{\Gamma},\tau_{\Gamma^\loc}} \; \sum_{\alpha_{\lbrace \Gamma \rbrace}} A^{[l+1]}(\Gamma,\Gamma^\loc,\Gamma')_{t_{\Gamma} \, \tau_{\Gamma^\loc} \, \alpha_{\lbrace \Gamma \rbrace}}^{t_{\Gamma'}} \times \nonumber \\ 
  & & \sum_{m_\Gamma, \mu_{\Gamma^\loc}} C(\Gamma,\Gamma^\loc,\Gamma')_{m_\Gamma \, \mu_{\Gamma^\loc}}^{m_{\Gamma'} \, \alpha_{\lbrace \Gamma \rbrace}} 
  \ket{\Gamma; t_{\Gamma}, m_{\Gamma}}_{l} \otimes \ket{\Gamma^\loc; \tau_{\Gamma^\loc}, \mu_{\Gamma^\loc}}_{l+1} \; .
\end{eqnarray}
\end{widetext}
Here, to emphasize their tensor character,  the usual Clebsch-Gordan coefficients have been denoted in a somewhat unusual way, 
$ \cgcoeff{\Gamma,m_{\Gamma};\Gamma^\loc,\mu_{\Gamma^\loc}}{\Gamma',m_{\Gamma'}}_{\alpha_{\lbrace \Gamma \rbrace}}\to C(\Gamma,\Gamma^\loc,\Gamma')_{m_\Gamma \, \mu_{\Gamma^\loc}}^{m_{\Gamma'} \, \alpha_{\lbrace \Gamma \rbrace}} $, with $\alpha$ the so-called outer multiplicity label.\cite{Cornwell_vol2}
This label is usually introduced for more complex groups 
such as SU($n>2$) or cubic symmetries, e.g., where certain irreducible representations occur multiple times
in the product of two other representations.
The outer multiplicity  label is usually dropped for  symmetries such as O(3) or SU(2), 
but it proves extremely useful   to keep it even in these simple cases. 
The interpretation of Eq.~ \eqref{eq:NAMPS_recursion} is simple:
we use  Clebsch-Gordan coefficients to construct the 
multiplets $ \Gamma'$ from  representations $ \Gamma $ and $ \Gamma^\loc $, and then mix these with the  NA-tensor $A(\Gamma,\Gamma^\loc,\Gamma')$ to obtain the appropriate Schmidt state.
In the construction above, we have tacitly assumed that  the state $\ket {\Psi} $  is a 'singlet', i.e., 
that it transforms according to the trivial representation,  $ \Gamma = 0 $. Then the trivial symmetry structure
 of $\ket {\Psi} $   ensures that  Schmidt-states form multiplets. The construction can, however, 
be easily generalized to the case  $\Gamma\ne0$ (see Appendix~\ref{appendix_sec:Nontrivial_rep} for details).
\begin{figure}[t]
 \begin{center}
  \includegraphics[width=0.60\columnwidth]{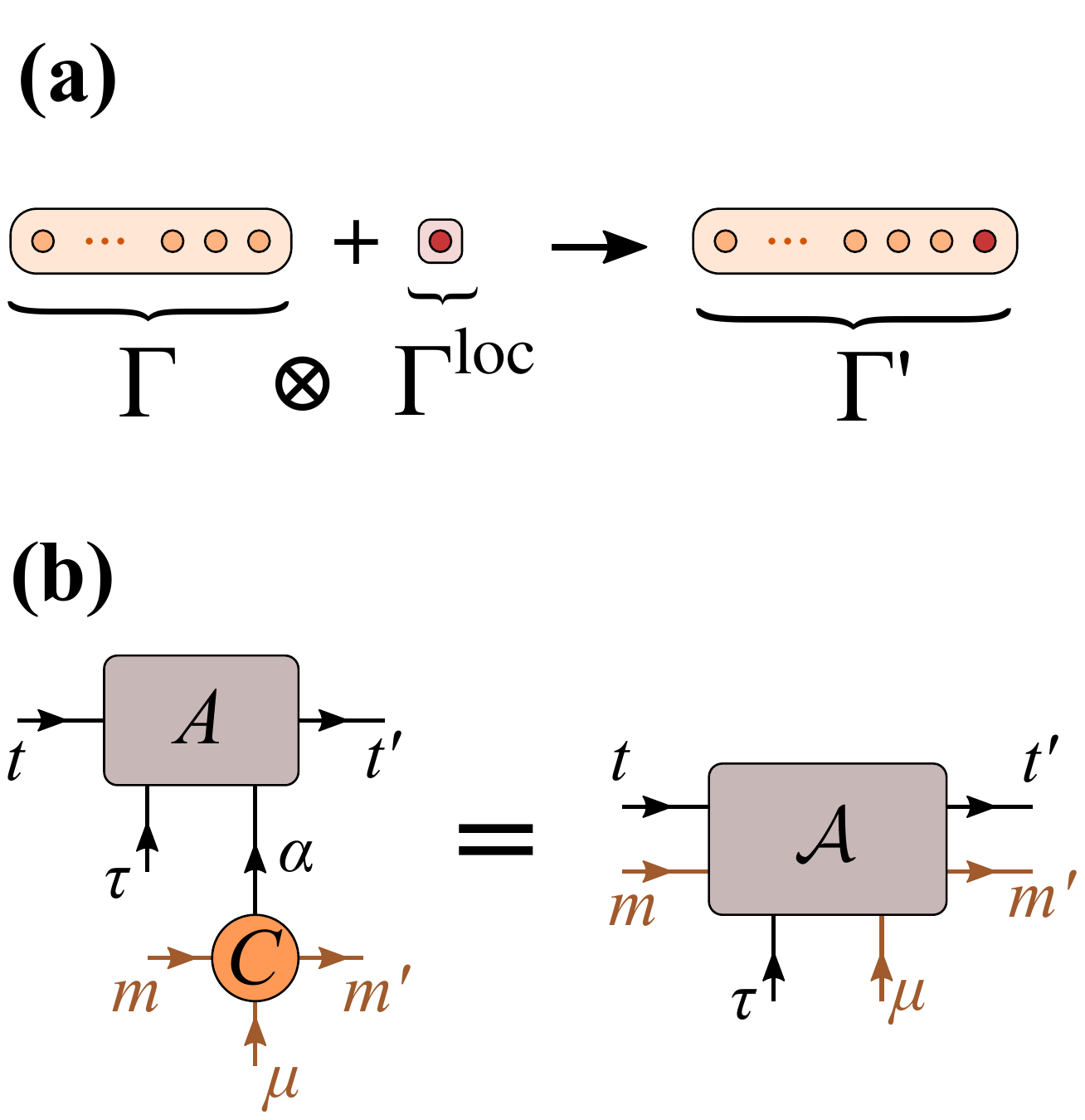}
 \caption{ Adding a site  to the left subsystem. (a) Schmidt states of the new, 'larger' subsystem  form multiplets  classified by the irreducible representations ($\Gamma'$). (b)  Corresponding tensor diagram, representing  Eqs.~\eqref{eq:NAMPS_rotation_light} and \eqref{eq:sym_nonsym_MPS_connection}. 
  Contracting the label $ \alpha $  yields a standard  MPS representation, Eq.~\eqref{eq:left_MPS_nonsym_full},
 which does not exploit symmetries.}
 \label{fig:add_one_site}
 \end{center}
\end{figure}

We pose here for a moment to  investigate the structure of the tensors appearing in the construction above.
The Clebsch-Gordan coefficient and the matrix $A$ are both  four-leg tensors, ogranized into blocks according to the three 
representation labels,  $ \lbrace \Gamma \rbrace = (\Gamma, \Gamma^\loc, \Gamma') $.  The external legs of these 
tensors are, however, tied to the block's considered. Certain legs, such as the $t$'s, $\tau$, the $m$'s and $\mu$,
depend only  on a particular representation, which we displayed as a label. The outer multiplicity label 
$\alpha$, however, depends on \emph{all three} $\Gamma$'s. These \emph{dependencies} play a crucial role 
in what follows: as we shall see, only  tensor legs with identical dependencies can be contracted. This is already clear in Eq.~
\eqref{eq:NAMPS_recursion}, where summation over the multiplicity label $\alpha$
\emph{enforces} the symmetry labels of  $A^{[l]}$ and $C$ to be identical. 
 
Eq.~\eqref{eq:NAMPS_recursion} is graphically represented in Fig.~\ref{fig:add_one_site}. 
We can rewrite 
 Eq.~\eqref{eq:NAMPS_recursion}  by simply suppressing the (quite obvious) dependency of the legs  as
\begin{widetext}
\begin{equation}\label{eq:NAMPS_rotation_light}
 \ket{\Gamma'; t', m'}_{l+1} = \sum_{\Gamma,\Gamma^\loc} \; \sum_{t,\tau} \; \sum_{\alpha} A^{[l+1]}(\lbrace \Gamma \rbrace)_{t \, \tau \, \alpha}^{t'} \sum_{m, \mu} C(\lbrace \Gamma \rbrace)_{m \, \mu}^{m' \, \alpha} 
  \ket{\Gamma; t, m}_l \otimes \ket{\Gamma^\loc; \tau, \mu}_{l+1} \; 
\end{equation}
The direct relationship between Eq.~\eqref{eq:NAMPS_rotation_light} and  \eqref{eq:recursion} 
can be established by summing over the outer multiplicity label  $ \alpha $ {(see also Fig.~\ref{fig:add_one_site}.b)},
\begin{equation}
\label{eq:sym_nonsym_MPS_connection}
 \sum_{\alpha} A^{[l+1]}(\brc{\Gamma})_{t \, \tau \, \alpha}^{t'} C(\brc{\Gamma})_{m \, \mu}^{m' \, \alpha} = \mathcal{A}^{[l+1]}(\lbrace \Gamma \rbrace)_{t \, m \; \;  \tau \,\mu}^{t'\, m'} \; .
\end{equation}
Iterating Eq.~\eqref{eq:NAMPS_rotation_light}, we arrive at the  
left-canonical NA-MPS representation of the state $\ket {\Psi}$,
\begin{eqnarray}\label{eq:NAMPS}
\ket{\Psi} = \sum_{\brc{\Gamma^{\loc}_l}} \sum_{\brc{\Gamma_l}} \sum_{\brc{ t_l}} \sum_{\brc{\tau_l}} \; \sum_{\brc{\alpha_l}} & & A^{[1]}(\brc{\Gamma}^{[1]})_{\tau_1 \, \alpha_1}^{t_1} \,  A^{[2]}(\brc{\Gamma}^{[2]})_{t_1 \, \tau_2 \, \alpha_2}^{t_2} \, \dots \,  A^{[L]}(\brc{\Gamma}^{[L]})_{t_{L-1} \, \tau_L \, \alpha_L} \nonumber \\ 
 \sum_{\brc{ m_l}} \sum_{\brc{\mu_l}} & & C(\brc{\Gamma}^{[1]})_{0 \, \mu_1}^{m_1 \, \alpha_1} \, 
C(\brc{\Gamma}^{[2]})_{m_1 \, \mu_2}^{m_2 \, \alpha_2} \, \dots \, C(\brc{\Gamma}^{[L]})_{m_{L-1} \, \mu_L}^{0 \, \alpha_L} \nonumber \\ & &  \ket{\Gamma^\loc_1; \tau_1,\mu_1} \otimes \ket{\Gamma^\loc_2; \tau_2,\mu_2} \otimes  \dots \otimes \ket{\Gamma^\loc_L; \tau_L,\mu_L} \; ,
\end{eqnarray}
\end{widetext}
where we have added a site label $l$ to the general notation $  (\Gamma_{l-1}, \Gamma^{\loc}_l, \Gamma_l)  \to \brc {\Gamma}^{[l]}$,  with $\Gamma_l$  denoting the representation indices of Schmidt states on the left of bond $l$. The formally introduced  representation index $ \Gamma_0 =0 $ stands for the 'empty' site, $l=0$, while 
$ \Gamma_L = 0$ is just the trivial representation to which the state $\ket{\Psi}$ belongs. 
(Generalization to $\Gamma_L \ne 0$ states is discussed in Appendix \ref{appendix_sec:Nontrivial_rep}.) 

  \begin{figure}[t]
\begin{center}
 \includegraphics[width=0.9\columnwidth]{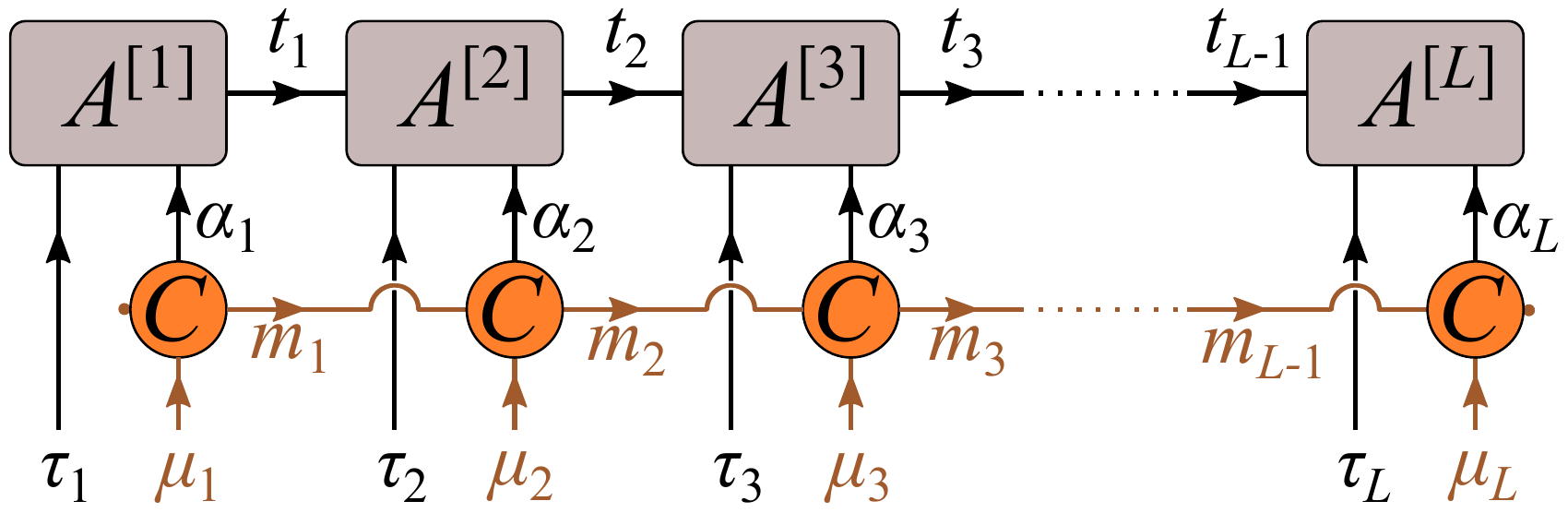}
 \caption{Representation of the NA-MPS in equation~\eqref{eq:NAMPS}. Multiplet indices $t_l, \tau_l $ and the outer multiplicity indices $ \alpha_l $ are shown in black, while  internal $ \mu_l, m_l $ indices of the representations are marked 
 by orange. The upper layer is free from internal indices of representations, but 
 has a block structure, as classified  by the  representation labels $\{\Gamma\}$.  
Clebsch-Gordan coefficients form  the lower layer.
 }\label{fig:NAMPS}
\end{center}
 \end{figure}
 
  Fig.~\ref{fig:NAMPS} shows a graphical representation for the NA-MPS state in Eq.~\eqref {eq:NAMPS}. It is constructed  as a two layer structure, with the lower layer containing  symmetry/representation-specific  information, encoded  through Clebsch-Gordan coefficients. The upper layer has, of course, also some knowledge about the underlying symmetry, since its blocks are labeled by the irreducible representations, but does not contain representation-specific information.
This two-layer structure is somewhat  similar to those in the  SU(2) symmetric implementations presented in Refs.~\onlinecite{SinghVidal2010,Gunst2019}. Here, however, we  take also keep track of  outer multiplicities in a very general way, which allows us to treat symmetries beyond SU(2) in a \emph{unified}, transparent,  and \emph{symmetry group independent} manner.  
 
Our goal is to eliminate the lower layer, and perform  DMRG or TEBD   \emph{only} on the upper layer, which can thus be considered as a full-fledged representation of the state $\ket{\Psi}$.
Removing the Clebsch layer   improves efficiency  in two ways:
 {\it i)} Since bond indices $t_l$ in the upper layer stand for multiplets instead of states, the bond dimension $M_{\mathrm{mult}}$ of the upper layer corresponds to a much larger conventional ('non symmetric') bond dimension.
{\it ii)} The representation indices $ \brc {\Gamma}^{[l]} $ must respect symmetry-specific  selection rules.
These selection rules allow for a very efficient {sparse block storage}, tensor multiplication, and  singular value decomposition (SVD).\cite{SVD_reference}

Although we refer to  the tensor structure introduced as non-Abelian, it naturally  incorporates 
familiar Abelian symmetries, too. For Abelian symmetries such  $U(1)$ or parity (${\cal Z}_2$), e.g.,  
 all representations are one-dimensional,   all  'Clebsches'  are just ones {for blocks allowed by the selection rules}, and the regular MPS structure is recovered {with efficient sparse block tensors in the decomposition}.

\section{NA-tensors}\label{sec:NA-tensor}

\subsubsection{NA-tensors and dependencies}

The  tensors $ A^{[l]} $, and the Clebsch-Gordan coefficients $ C $ {in Eqs.~\eqref{eq:NAMPS_rotation_light} and \eqref{eq:NAMPS}}, have 
the same fundamental structure, which we  refer to as  
\emph{non-Abelian tensor} (NA-tensor).  
General  NA-tensors $T (\brc {\Gamma})_{i_1 \, i_2 \dots i_n} ^ {j_1\, j_2 \dots j_m} $, 
have a structure shown  in Fig.~\ref{fig:NAtensor}: 
they have a block structure  with blocks labeled  by lists of symmetry labels  (representation indices),\
 $\brc {\Gamma} = (\Gamma_1, \dots, \Gamma_k) $, with each $\Gamma_i$ referring to 
 a list of quantum numbers used.\footnote{In case of multiple symmetries like $SU(3) \times U(1)$ each label $\Gamma$ is alone a combination of two quantum numbers, specifying an SU(3) representation (Young tableau) and a U(1) charge. Then a block of an NA-tensor is addressed by a set of such combined representation labels.} 
  They have, furthermore, external incoming and outgoing legs. 
 Since many blocks contain only zeros by selection rules, 
an  efficient sparse block storage can be achieved by storing only non-zero blocks. 
 Block sizes {usually} depend on the {specific set of representations} $\brc {\Gamma}$, 
 and can have different block sizes at every site.

The legs of  NA-tensors have  implicit dependence on the tensor's  internal
symmetry labels. The label $t$ in Eq.~\eqref{eq:NAMPS_rotation_light},
 e.g., runs over multiplets belonging to a given representation, $\Gamma$. 
Similarly,  the index   $ m $   can take $ \dim (\Gamma) $ different values.
Thus both  $ t $ and $ m $ \emph{depend} on the representation index $ \Gamma $.
The outer multiplicity index $ \alpha $  depends on 
all three  representation indices labeling a given (non-vanishing) symmetry block of the {Clebsch-Gordan tensor}, as well as 
that of the tensor $A^{[l]}$.  Generally, any given leg of an NA-tensor depends on a given 
subset of the representations $\brc {\Gamma}$, labeling  the blocks.

\subsubsection{Multiplication rules}\label{subsec:NAtensor_dot}
 The {MPS representation  in} Eq.~\eqref{eq:NAMPS} (see also Fig.~\ref {fig:NAMPS}) 
 allows us to introduce multiplication rules. By construction,
 the tensors $ A^{[l]} $  and $ C$ belonging to the same site are glued together such that 
 the  three representation indices $\brc{\Gamma}^{[l]} = (\Gamma_{l-1}, \Gamma^\loc_{l}, \Gamma_{l})$
 are always the same. This is, in fact, \emph{enforced} by the {contraction of the} outer-multiplicity index  $ \alpha $. 
 Similarly, we may notice that only those  tensor blocks of  $ A^{[l]} $ and $ A^{[l + 1]} $ are 
 contracted through the index $t_l $, where the corresponding representation $\Gamma_l$  (associated with the bond 
 between sites $l$ and $l+1$) is identical. Similar observations can be made by investigating the 
 Clebsch-Gordan tensors. 
 
These observations lead us to the general  (graphical) 
contraction rule:
\begin{enumerate}
 \item[i)]Incoming legs of NA-tensors can be contracted with outgoing legs provided that {all their dependencies match.} 
\item[ ii)] The resulting tensor's blocks are labeled by the representation tensor indices of the original tensors, but the matched representation indices are listed only once.
 \end{enumerate}

In large tensor networks, such as {the ones displayed in} in  Figs.~\ref{fig:NAMPS_dot} and \ref{fig:NAMPS_scalarop_matrixelement}, 
virtually all  legs are contracted, but  the rules above would result in NA-tensors whose blocks are still labeled by all  representation indices $ \Gamma^\loc_l $ and $ \Gamma_l $, while most of the representation indices are 
redundant in the sense that remaining legs do not depend on them.  Note also that  Eq.~\eqref {eq:NAMPS} contains a summation over  representations index sets, $ \brc {\Gamma_l} $ and $ \brc {\Gamma^\loc_l} $. It is therefore useful to introduce  the following  rule,
\begin{enumerate}
 \item[iii)]
{If there is one or more representation indices in the result tensor that no remaining (uncontracted) legs depend on, then  blocks must be summed over these representation indices.}
\end{enumerate}
This rule eliminates redundant representation indices. 

{At the end of this section, let us compare our NA-tensors with the 'QSpaces' tensors introduced by Andreas Weichselbaum.\cite{Weichselbaum2012, Weichselbaum2020} The main difference between the two approaches is the handling of Clebsch-Gordan coefficients. In Ref.~\onlinecite{Weichselbaum2012}  tensors are more complicated objects: they have not just blocks, but every block consists of more layers: one layer contains the representation-independent parts of the tensor (this layer correspond to our $A^{[l]}$ tensors), while other layers contain the Clebsch-Gordan coefficients (or their various combinations) for different symmetries. In these abstract tensors  structural and Clebsch-Gordan blocks are 
{grouped together} for every enabled set of representation labels. This multilayer structure leads to sophisticated 
 multiplication rules.\cite{Weichselbaum2020} In contrast, in our approach we \emph{ separate completely} 
 the Clebsch-Gordan coefficients from the structural $A^{[l]}$ tensors, and collect them into the $C$ tensor, whose mathematical structure is essentially the same as that of the $A^{[l]}$ tensors. As a result, our NA-tensors are conceptually simpler objects with relatively simple multiplication rules.

\section{Time evolving block decimation with NA-tensors }\label{sec:NATEBD}

\subsection{Basic steps of TEBD}
\label{sec:TEBD_nonsym}

We now demonstrate the NA-MPS approach on one of the simplest MPS algorithms, the
  time evolving block decimation (TEBD). This method, originally
 introduced by Guifr\'e Vidal,\cite {Vidal2004, Vidal2007} has since been 
exhaustively used to simulate one dimensional quantum systems in out of equilibrium.\cite{Znidaric_2008,Bardarson_2012,Pozsi_and_us_2014,Schreiber_2015,Else_2016}  
We consider here  Hamiltonians  with nearest neighbor interactions,
\begin{equation}\label{eq:Hamiltonian_nearest_neighbour}
 \hat{H} = \sum_{i=1}^{L-1} \hat{h}^{(2)}_{i,i+1} \; ,
\end{equation}
with $\hat{h}^{(2)}_{i,i+1}$ acting on sites $i$ and $i+1$.  Within TEBD, one divides $\hat H $
into parts acting on even and odd bonds,
\begin{equation}
 \hat{H} = \hat{H}_{\mathrm{even}} + \hat{H}_{\mathrm{odd}} = \sum_{k} \hat{h}^{(2)}_{2k,2k+1} + \sum_{k} \hat{h}^{(2)}_{2k+1,2k+2} \; , 
\end{equation}
and 'Trotterizes' the time evolution operator $e^{-i\hat H t}$, i.e., divides time into small segments of length $\Delta t$, 
and then applies a second order Trotter-Suzuki approximation~\cite {Trotter, Suzuki}, 
$e^{-i\Delta t \hat H}\approx e^{-i\Delta t \,\hat H_{\mathrm{even}}/2}  e^{-i\Delta t \,\hat H_{\mathrm{odd}}}  e^{-i\Delta t \,\hat H_{\mathrm{even}}/2}$. This procedure yields the time evolution, represented in Fig.~\ref{fig:TEBD_SuzukiTrotter}. 
Time evolution occurs on bonds, and after each step, singular value decomposition (SVD) can be used to reconstruct the 
original MPS structure of the state $\ket{\Psi(t)}$.

\begin{figure}[t]
 \includegraphics[width=1.0\columnwidth]{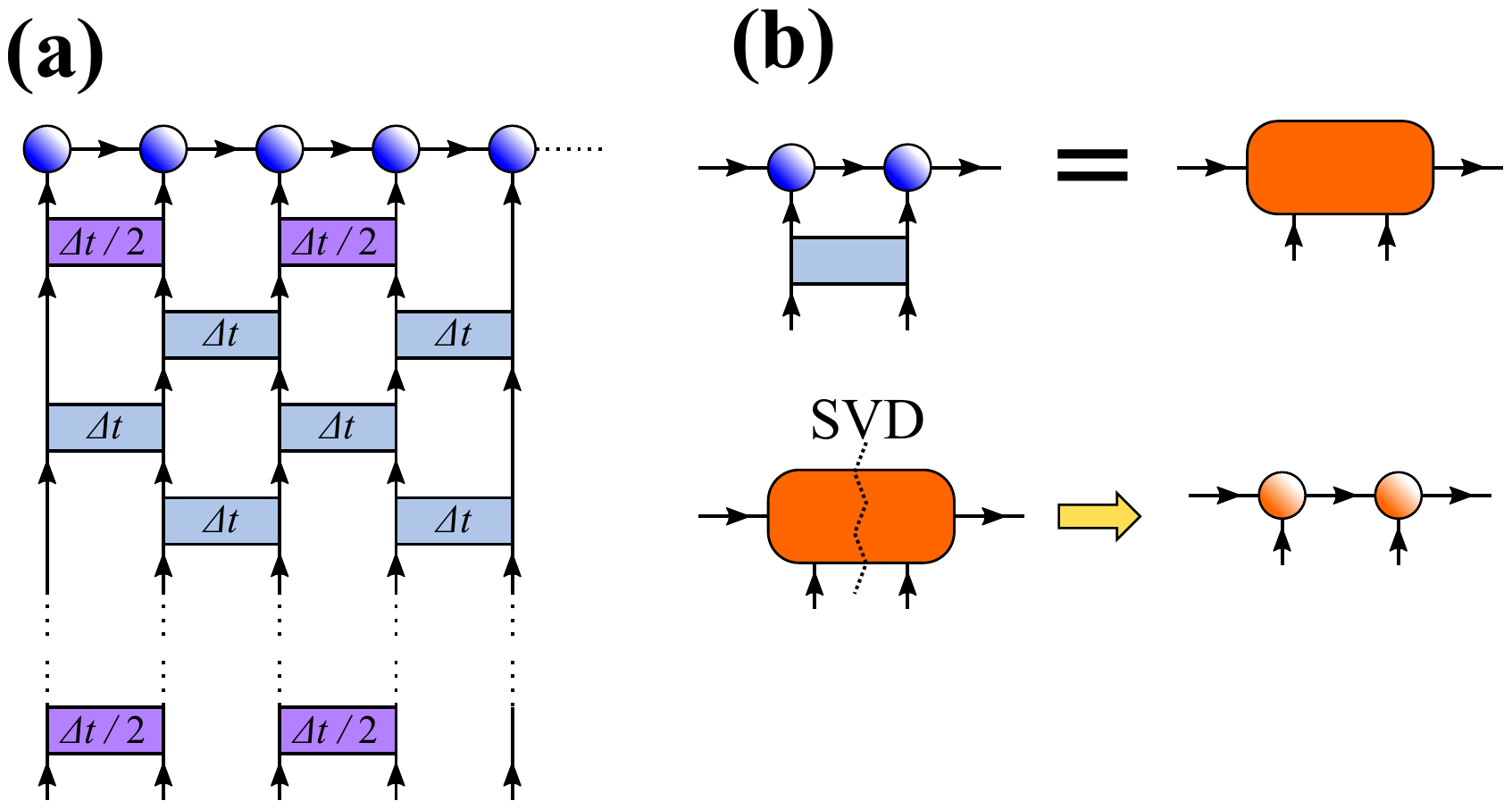}
 \caption{(a) Time evolution with second order Trotter-Suzuki approximation, yielding 
 a sequence of  two-site operations,  generated by the even and odd bond parts of 
 the Hamiltonian. 
 (b) SVD is used after each  step to restore the original  MPS structure.} 
 \label{fig:TEBD_SuzukiTrotter}
 \end{figure}

\subsection{TEBD with NA-MPS }\label{subsec:NATEBD}
\begin{figure}
\begin{center}
 \includegraphics[width=1.0\columnwidth]{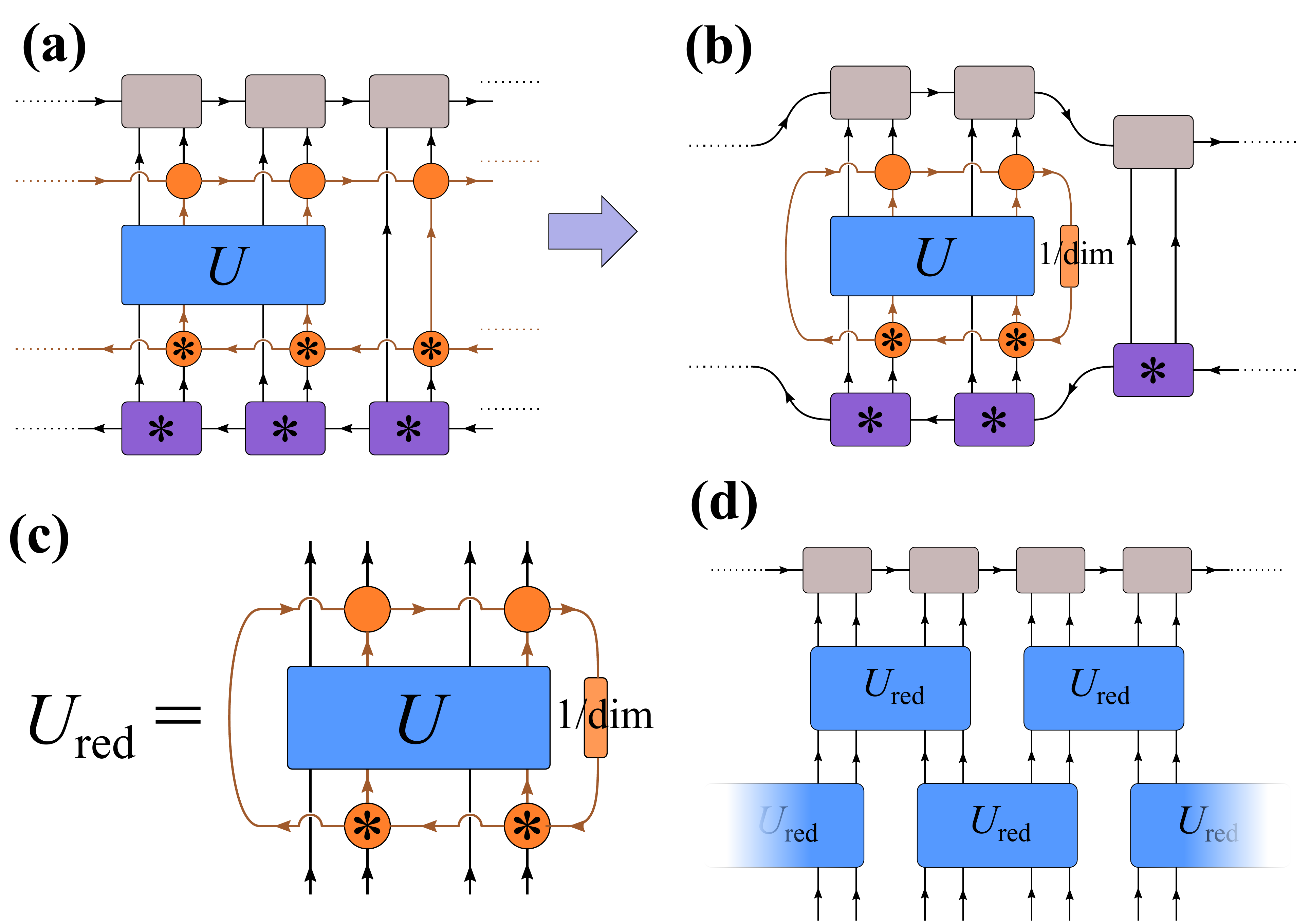}
 \caption{(a)  Matrix element of the time evolution operator  between two  NA-MPSs. (b-c)  Almost all Clebsch-Gordan tensors 
 can be eliminated using orthogonality relations, leading to a reduced time evolution operator. (d)
 The reduced tensors $ U_\mathrm{red} $  act then directly on  the top layer of the NA-MPS.} \label{fig:NATEBD}
\end{center}
\end{figure}

We now extend TEBD to  non-Abelian MPS's to obtain the non-Abelian version of TEBD (NA-TEBD). 
Here we focus on the key steps and use a graphical language {(see Fig.~\ref{fig:NATEBD})}. Technical details 
 are relegated  to  Appendices~\ref{appendix_sec:NAMPS_states} and \ref{appendix_sec:NATEBD}. 

The crucial step is the construction of a  \emph{reduced 
evolution operator} $U _{\mathrm {red}}$, which   incorporates unnecessary  Clebsch-Gordon coefficients,   
and time evolves  only  the upper layer of the NA-MPS state. 
The overall construction of the $U _{\mathrm {red}}$
is presented in Fig.~\ref{fig:NATEBD}.  To obtain  $U _{\mathrm {red}}$, we 
 compute  the overlap $\bra{\tilde{\Psi}} U \ket{\Psi}$,  with both states being written in the NA-MPS form (such an overlap is graphically displayed in panel (a) in Fig.~\ref{fig:NATEBD}). Using the orthogonality properties of 'Clebsches' (see 
Appendix~\ref{appendix_sec:NAMPS_states}), we can eliminate all but four Clebsch-Gordan tensors, which are then incorporated into the reduced evolution operator (panel (b) in Fig.~\ref{fig:NATEBD}). 
This leaves us with the reduced two-site  propagator, 
 $U_{\mathrm {red}} (\brc  {\Gamma})_{\tau_l'\, \alpha_l'\; \; \tau_{l + 1} '\, \alpha_ {l + 1}'} ^ {\tau_l \, \alpha_l \; \;  \tau_{l + 1} \, \alpha_{l + 1}} $, acting on sites $l$ and $l+1$ (panel (c) in Fig.~\ref{fig:NATEBD}).

 Notice that this eight-leg NA-tensor  is labeled by  a total of eight representation indices: 
$ \brc {\Gamma} = \left (\Gamma_{l-1}, \Gamma_l^\loc, {\Gamma_l^\loc} ', \Gamma_l, \Gamma_l', \Gamma_{l + 1}^\loc, {\Gamma_{l + 1}^\loc}', 
\Gamma_{l + 1} \right) $. 
 The indices $\tau$ depend just on    local representations, while the {dependencies of} the 
 outer multiplicities  $ \alpha $  read  
 $ \mathrm {dep} (\alpha_l) = (\Gamma_{l-1}, \Gamma_l^\loc , \Gamma_l) $, 
$ \mathrm {dep} (\alpha_{l + 1}) = (\Gamma_{l}, \Gamma_{l + 1}^\loc, \Gamma_{l + 1}) $ , 
$\mathrm {dep} (\alpha_l ') = (\Gamma_{l-1}, {\Gamma_l^\loc}', \Gamma_l ') $,
 and $ \mathrm {dep} (\alpha_{l + 1}') = (\Gamma_l', {\Gamma_{l + 1}^\loc}', \Gamma_{l + 1}) $.
 Notice that, numerically, it is sufficient to compute the reduced evolution operator $U _{\mathrm {red}}$ 
only once. 

Having discarded the 
Clebsch-layer, the reduced operator  now acts only on the upper layer of the $ A^{[l]} $-tensors. From now on, there is no  significant difference between the NA-TEBD and the usual TEBD;
the upper layer of NA-MPS behaves in the simulations like  a normal 
MPS  that is updated at each time step, only the singular value decomposition step,
discussed in more detail in Appendix~\ref{appendix_sec:NATEBD}
 requires some care {(see also Fig.~\ref{fig:NATEBD} (d))}.  

Since NA-TEBD  is formulated in terms of  the upper layer of the NA-MPS, one does not need to take 
care of {internal states of multiplets},   and necessary numerical  resources   are  determined   by the  bond dimension of 
the upper layer, $ M_{\mathrm {mult}}$. Due to the block structure of the $ A^{[l]} $-tensors, the SVD transformation 
can be performed separately according to the  representation indices of the Schmidt states. 
In this way, we can reach bond dimensions {in the range of 
tens of thousands} in terms of  usual, non-symmetric or Abelian states {even on simple desktop computers.}

\section{Application to the  SU(3) Hubbard model}
\label{sec:Hubbard}
We now illustrate the advantages of  NA-TEBD by simulating a quantum quench
 on the  one dimensional Hubbard model, Eq.~\eqref{eq:SU3},   at $1\over 3$-filling.

In this case, the local Hilbert space is $d=2^3=8$ dimensional. The model defined by  Eq.~\eqref{eq:SU3} 
possesses a U(1) charge symmetry, generated by the total charge, 
$$
\hat Q \equiv \sum_l \;\hat q_l\,,\quad \quad\mbox{with}\quad \hat q_l =  \sum_{\alpha} (c^\dagger_{l,\alpha} c_{l,\alpha} -1/2)\;,
$$
and an SU(3) symmetry generated by the eight SU(3) generators, 
$$
\hat \Lambda^i \equiv \sum_l \; {\hat \lambda}^i_l\;, \quad \quad\mbox{with}\quad \hat \lambda^i_l =  \sum_{\alpha,\beta} c^\dagger_{l,\alpha} \lambda^i_{\alpha\beta} c_{l,\beta}\;.
$$
Here the $\lambda^i$ denote the usual Gell-Mann matrices, satisfying the  SU(3) Lie algebra, $[\lambda^i,\lambda^j]
= i f^{ijk}\lambda^k$. The Hamiltonian commutes with all generators above, and 
has a corresponding  $ \mbox{SU(3)} \times \mbox{U(1)} $  symmetry. 

The local Hilbert space at each site  is spanned by  four
multiplets, organized according to the total charge $Q^\loc$ and an $SU(3)$ representation label, 
typically specified by a Young tableau  (see Table~\ref{tab:fermion_site}). 
In case of SU(3), possible Young tableaux consist if two lines, and the length of these 
lines $F\equiv (m_1,m_2)$ specify the representation.\footnote{This is different from SU(2), where every Young tableau
consists of a single line of length $2S$, with $S$  the usual spin.} The local representation label 
is therefore a composite label,  $ \Gamma^\loc_l = \{F^ \loc, Q^\loc\} $.   

\begin{table}[b]
\begin{tabular}{|c| c| c| c|}
 \hline
 $\Gamma^{\loc} = (F^\loc,\; Q^\loc) $ &  $\dim(\Gamma^\loc)$ & $\dim(\tau)$ & states  \\
 \hline \hline
 (\textbullet,\; 0  ) & 1 & 1 & $\ket{0}$ \\ 
 \hline\
  &  &  & $c^\dag_{1} \ket{0}$\\  
 ( ${\yng(1)} $ ; 1) & 3 & 1 & $c^\dag_{2} \ket{0}$\\ 
  &  &  & $c^\dag_{3} \ket{0}$\\ 
  \hline\
  &  &  & $c^\dag_{1} c^\dag_{2} \ket{0}$\\   
 $\Big(\,\yng(1,1)\, ; 2\Big )$  & 3 & 1 &   $c^\dag_{2} c^\dag_{3} \ket{0}$ \\
 &  &  & $c^\dag_{3} c^\dag_{1} \ket{0}$\\
  \hline\   
 (\textbullet,\; 3  ) & 1 & 1 & $c^\dag_{1} c^\dag_{2} c^\dag_{3} \ket{0}$\\
 \hline
\end{tabular}
\caption{$SU(3)$  local states and representations. SU(3) representations are denoted by Young tableaux.}\label{tab:fermion_site}
\end{table}

We start our simulations from a state $|\Psi(0)\rangle_0$, where three particles are localized at every third site 
(see Fig.~\ref{fig:quench_sketch}), 
\begin{equation}
\ket{\Psi(0)}_0 = \prod_{\alpha=1}^3 \prod_{l=3k} c^\dagger_{l\alpha}\ket{0}\;.
\end{equation}
This state has clearly an MPS structure and is, moreover, an SU(3) singlet.

\subsection{Non-interacting time evolution}

 To test our NA-TEBD approach, we first consider time evolution in the case $U=0$. Then the problem is exactly solvable, and 
 we can compute all correlation functions and expectation values analytically. We just need to observe that for $U=0$,
 the time evolved  wave function $\ket{\Psi(t)}_0$ can be written as a Slater determinant, 
 \begin{equation}
\ket{\Psi(t)}_0 = \prod_{\alpha=1}^3 \prod_{l=3k} c^\dagger_{l\alpha}(t)\ket{0}\;,
\end{equation}
 with the time evolved operators expressed as
 \begin{equation}
 c^\dagger_{l\alpha}(t) = \int_{-\pi}^\pi \frac{{\rm d} p}{2\pi}\;
e^{-i \,l\cdot p}  e^{-i \,2J \cos(p) \,t} c_\alpha(p)\;.
\end{equation}
The occupation at site $l=0$ can then be expressed as 
\begin{equation}
n^{U=0}_0(t) = \bra{\Psi(t)} \sum_\alpha c^\dagger_{0,\alpha} c_{0,\alpha} \ket{\Psi(t)}_0\;.
\end{equation}
This expectation value can be evaluated by Wick's theorem, yielding 
\begin{equation}
n_0^{U=0}(t) =  \sum_{l = 3k}
\iint\limits_{-\pi}^{\quad\;\pi} \frac{{\rm d} p\, {\rm d} p'}{(2\pi)^2}  
e^{-i \,l\cdot (p-p')}  e^{-i\,2J\, t (\cos(p)-\cos(p'))} \;.
\end{equation}
Carrying out the summation over $l$ yields the $2\pi$ periodic delta function, 
$\delta_{2\pi}(3(p-p'))$, which can be used to eliminate one of the momentum integrals, finally yielding
\be
n^{U=0}_0(t) = 1+ 2\,J_0(2 \sqrt{3}\, J\,t)\;
\ee
for the non-interacting case, $U=0$, with $J_0$ the Bessel function of the first kind. 
The value of $n_1(t)$ follows simply from particle number conservation, 
\be
n^{U=0}_1(t) = 1- J_0(2 \sqrt{3}\, J\,t)\;.
\ee
Thus charge oscillations decay algebraically as $1/\sqrt{t}$ in the non-interacting case.

\begin{figure}[t]
 \begin{center}
  \includegraphics[width=1.0\columnwidth]{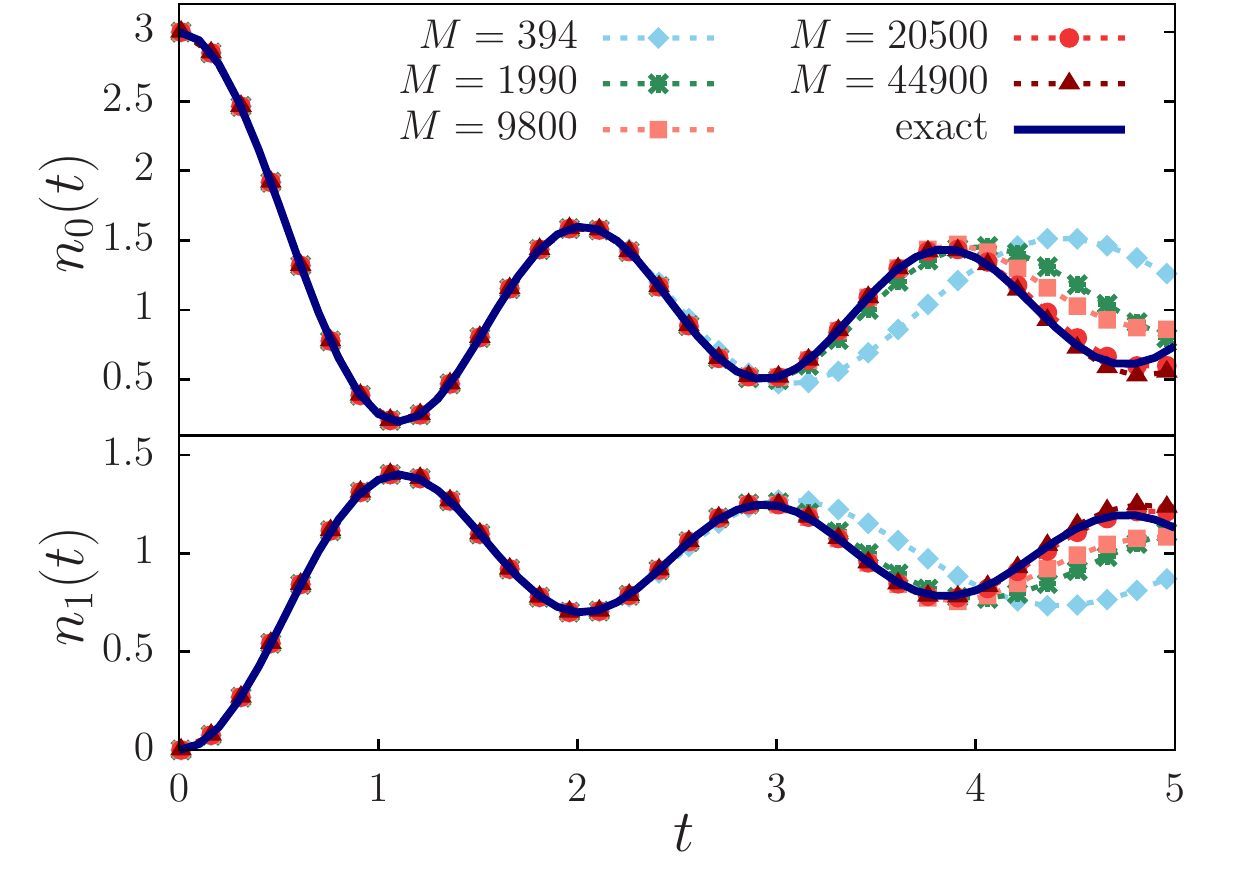}
 \caption{Charge oscillation at the origin, $n_0(t)$, and at the 
 first nearest neighbor, {$n_1(t)$}, in the absence of interactions, $U=0$, for various bond dimensions.}\label{fig:chargeU0}
 \end{center}
\end{figure}

As shown in  Fig.~\ref{fig:chargeU0}, this algebraic decay is well captured by NA-TEBD for short times, however, 
to capture the second oscillation, fairly large bond dimensions $\sim M\gtrsim 20,000$ 
are needed, corresponding to keeping $M_{\mathrm{mult}} = 2500$ multiplets. With NA-TEBD simulations, we can  easily reach these bond dimensions on a simple work station, 
which would be quite hopeless without exploiting the SU(3) symmetry. 

We can  test the accuracy of our computations also by investigating the 
increase of the bond entropy for $U=0$. We can compute this {latter} by using the approach of {Peschel and 
Eisler.\cite{Peschel2009}}
To compute the entanglement entropy in a non-interacting system, we  
consider a long enough segment $L$  of  the infinite one-dimensional system, and compute the correlator 
$C_{l,l'\in\ L }(t) \equiv \bra{\Psi(t)}c^{\dagger}_{l\alpha } c_{l'\alpha}\ket{\Psi(t)}_0$, 
which, for a non-interacting system contains all information on the 
reduced density operator. 
The correlator $C_{l,l'\in\ L }(t)$  can be evaluated along similar lines as 
the expectation value, $n_0(t)$, and is given by
\be
C^{U=0}_{ll'}(t) = \frac1 3 + \frac 1 3 \,e^{i \frac \pi 3 (l+l')} \big( 
 1+ 
e^{i \frac \pi 3 (l+l')} \big)\, J_{l-l'} (2 J t \sqrt{3})\;.
\ee
As shown in Ref.~\onlinecite{Peschel2009}, the entanglement entropy between the segment $L$  and the rest of the system 
 can expressed just in terms of the eigenvalues $\xi$ of  $C_{l,l'\in\ L }(t)$ as
\be 
S^L_{\rm vN}(t) = - 3 \sum_\xi \big(\xi\ln(\xi) + (1-\xi)\ln(1-\xi)\big)\;,
\ee
where the factor 3 is due to the SU(3) flavor degeneracy. For large enough segments, this is just twice the 
entanglement entropy of two halves of an infinite system, 
\be 
S_{\rm vN}(t) 
 = \frac 1 2 \lim_{L\to\infty}  S^L_{\rm vN}(t)\;.
\ee
Computing the eigenvalues $\xi$ numerically, we can thus determine the complete 
time dependence of the  entanglement entropy, $S_{\rm vN}(t) $.

\begin{figure}[t]
 \begin{center}
  \includegraphics[width=1.0\columnwidth]{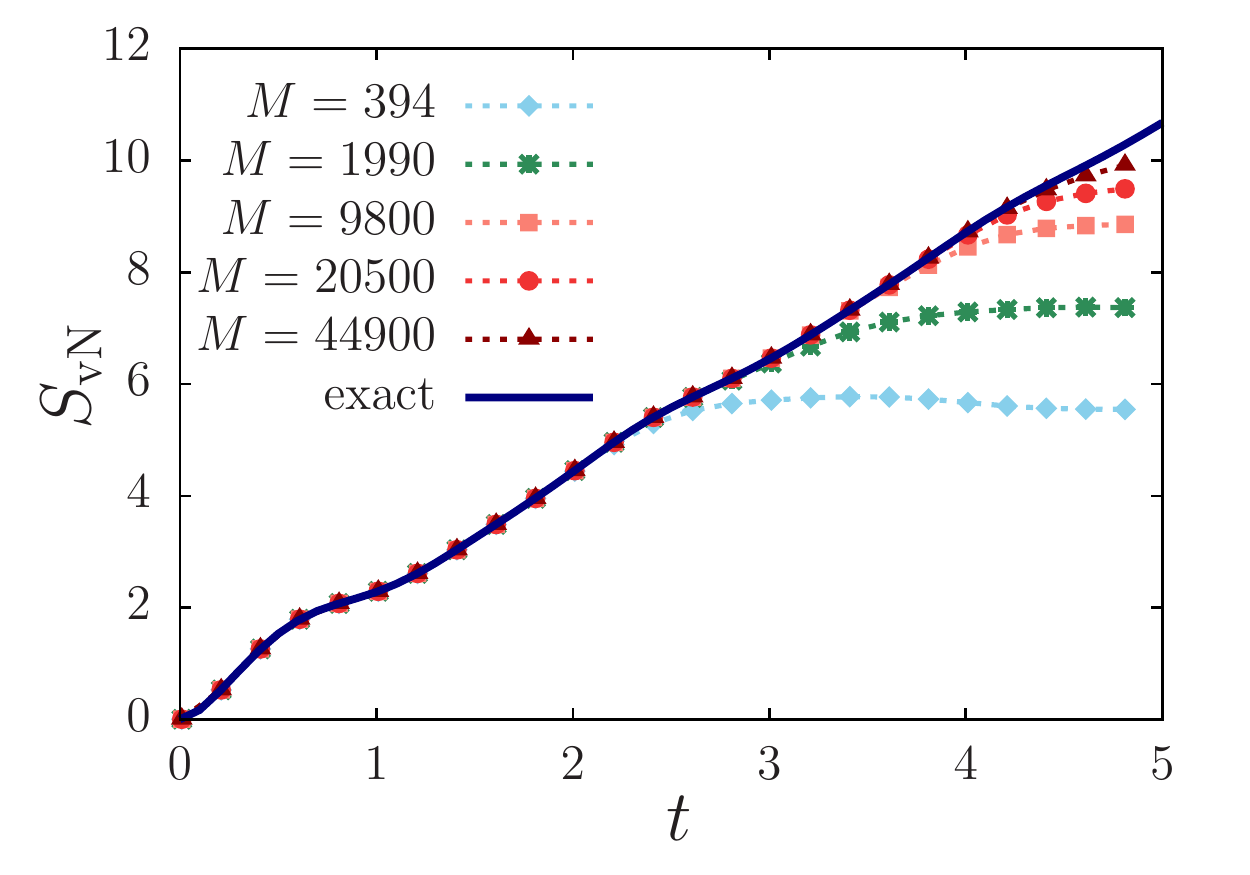}
 \caption{Entropy growth for $U=0$, as a function of bond dimensions. 
 Small oscillations are observed  on top of an overall  linear entropy growth. 
 Very large ($M\approx 45,000$) effective bond dimensions are needed to recover
 the exact results (blue line)  up to times $t\approx 4.5$. }\label{fig:entropyU0}
 \end{center}
\end{figure}

The (numerically determined) exact entanglement entropy is compared with the  NA-TEBD results 
in Fig.~\ref{fig:entropyU0}. The initial state is a product state, and therefore completely unentangled
at $t=0$. However, entanglement  is generated with time. 
The Neumann entropy starts to increase roughly linearly, as predicted for  gapless systems,~\cite{CalabreseCardy_entropygrowth}
but is modulated by small oscillations, reflecting the presence of coherent charge oscillations. 
NA-TEBD breaks down approximately where the bond dimension is  insufficient to 
keep track of the entanglement entropy. In this gapless system, the conformal central charge is 
quite large, $c=3$,  implying a fast increase of entanglement entropy. 
Indeed, numerical computations are  quite demanding in this model,  and 
bond dimensions in the  range of $M\sim 40-50,000$ are needed to reach time scales
  {$t\gtrsim 4J^{-1}$}.


\subsection{Interaction effects}

\begin{figure}[t]
\begin{center}
  \includegraphics[width=1.0\columnwidth]{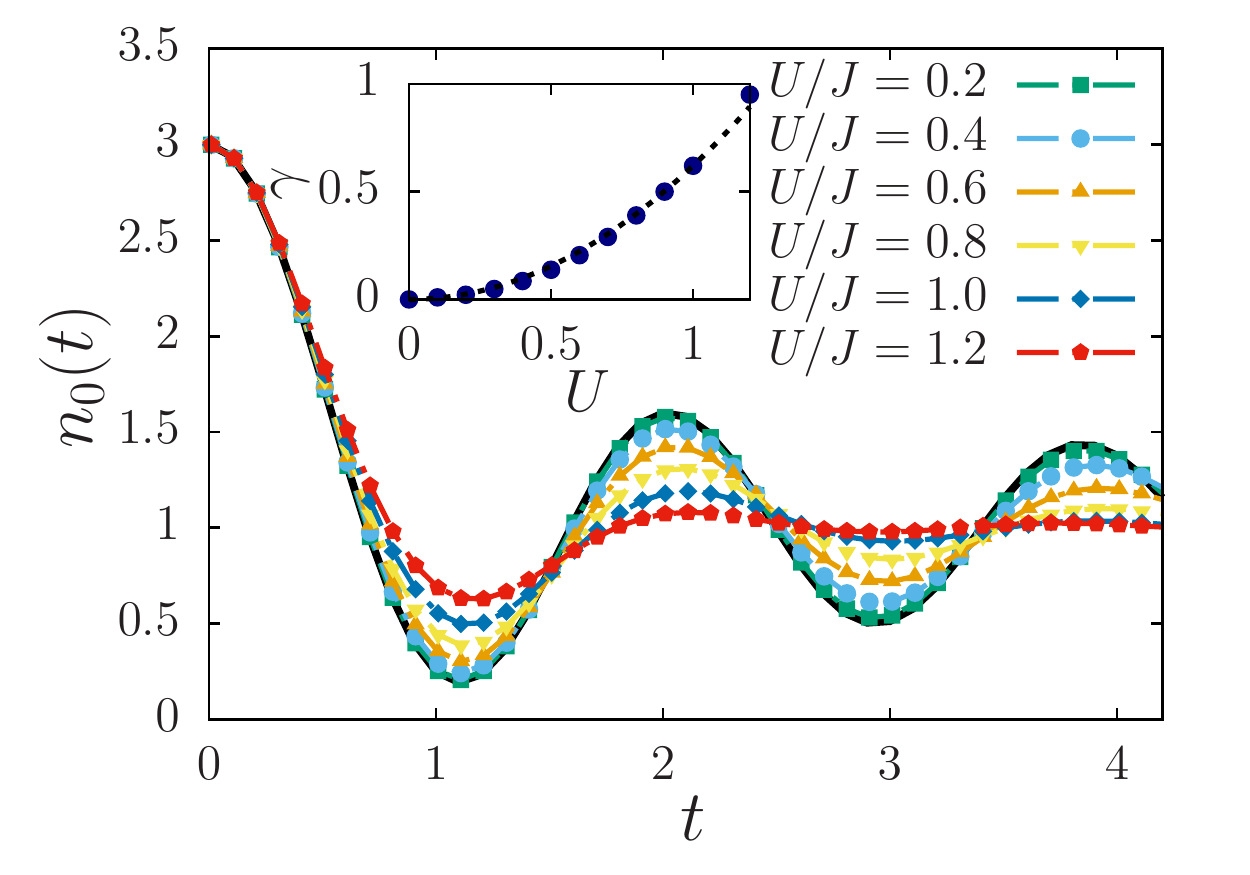}
 \caption{Charge on the initially triple-occupied site as a function of time for different interaction strengths $U$. 
 For small interaction strengths $U \lesssim 1.2$ we observe damped oscillations around the thermalized occupation $n = 1$.
 Inset: extracted damping rate as a function of $U$.}\label{fig:charge_smallU}
 \end{center}
\end{figure}

Interactions change the previous results dramatically. As shown in Fig.~\ref{fig:charge_smallU}, 
charge oscillations become rapidly damped with increasing $U$.   In the regime, $U\lesssim 1$,  charge oscillations 
{are suppressed}  exponentially in time compared to  free fermion oscillations, 
{$\delta n^{U\ne0}_0(t) \propto e^{-\gamma \, t} \cdot J_0(2 \sqrt{3}\, J\,t)$}.  The extracted 
damping rate $\gamma$ increases quadratically for small and moderate couplings, $U\lesssim 1 $, 
 as expected from perturbation theory, and  shown in the inset  of Fig.~\ref{fig:charge_smallU}.

 \begin{figure}[b]
 \begin{center}
  \includegraphics[width=0.9\columnwidth]{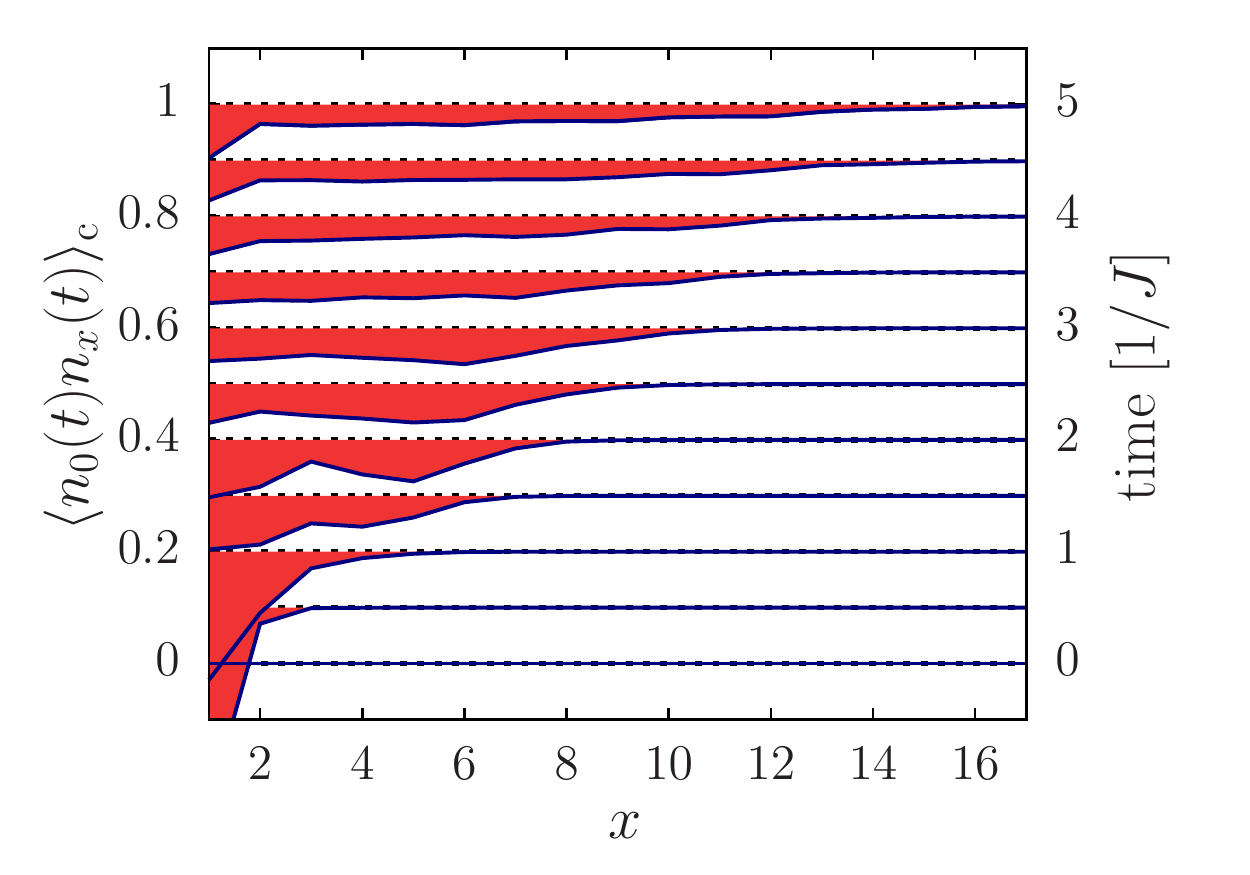}
 \caption{Connected part of charge-charge correlations computed for $U=1$. 
 Correlations spread relativistically with a velocity $v\sim J$ for small and moderate values of $U$, 
 but interactions quickly remove the oscillations and power-law correlations present in the non-interacting system.} 
 \label{fig:nn_corr}
 \end{center}
\end{figure}

 NA-TEBD can also be used  to compute time dependent correlation functions.
   The precise numerical 
procedure is outlined in Appendices~\ref{appendix_sec:NAMPS_states}
and  \ref{appendix_sec:Wigner_Eckart}. For {the} sake of simplicity, here we focus on 
the scalar operator, $\hat n_l$, which commutes with the symmetry generators, and is 
also completely local.  For such operators, we can easily construct the 'reduced operator', {$\hat O_l \to O_l(\Gamma^\loc_l)_{\tau}^{\tau'}$ 
(see Appendix~\ref{appendix_sec:NAMPS_states}), which acts directly on the the upper NA-MPS layer. }
From this point on, the computation of correlation functions
follows the same line as for Abelian symmetries or non-symmetrical MPS states.~\cite{Schollwoeck_Rev2010}

 Fig.~\ref{fig:nn_corr} shows the the time evolution of the connected correlator 
$$
C_{nn}^{\rm conn}(l, t) \equiv \langle \hat n_0(t) \hat n_l(t) \rangle  - \langle  \hat n_0(t) \rangle  \langle  \hat n_l(t) \rangle \,
$$
{for  interaction strength $U=1$}. The connected part of the correlation function is negative, 
indicating that excess particle densities 
emerge due to the quantum propagation  of  particles originally sitting at the origin. 
{The connected negative correlations trace a  light cone indicating that 
correlations and entanglement are both created by particles (or collectivemodes) traveling ballistically 
with a velocity $v\sim t$.}

\section{Numerical efficiency of NA-TEBD}
\label{sec:Numerical_test}

\begin{figure}[b]
 \begin{center}
  \includegraphics[width=1.0\columnwidth]{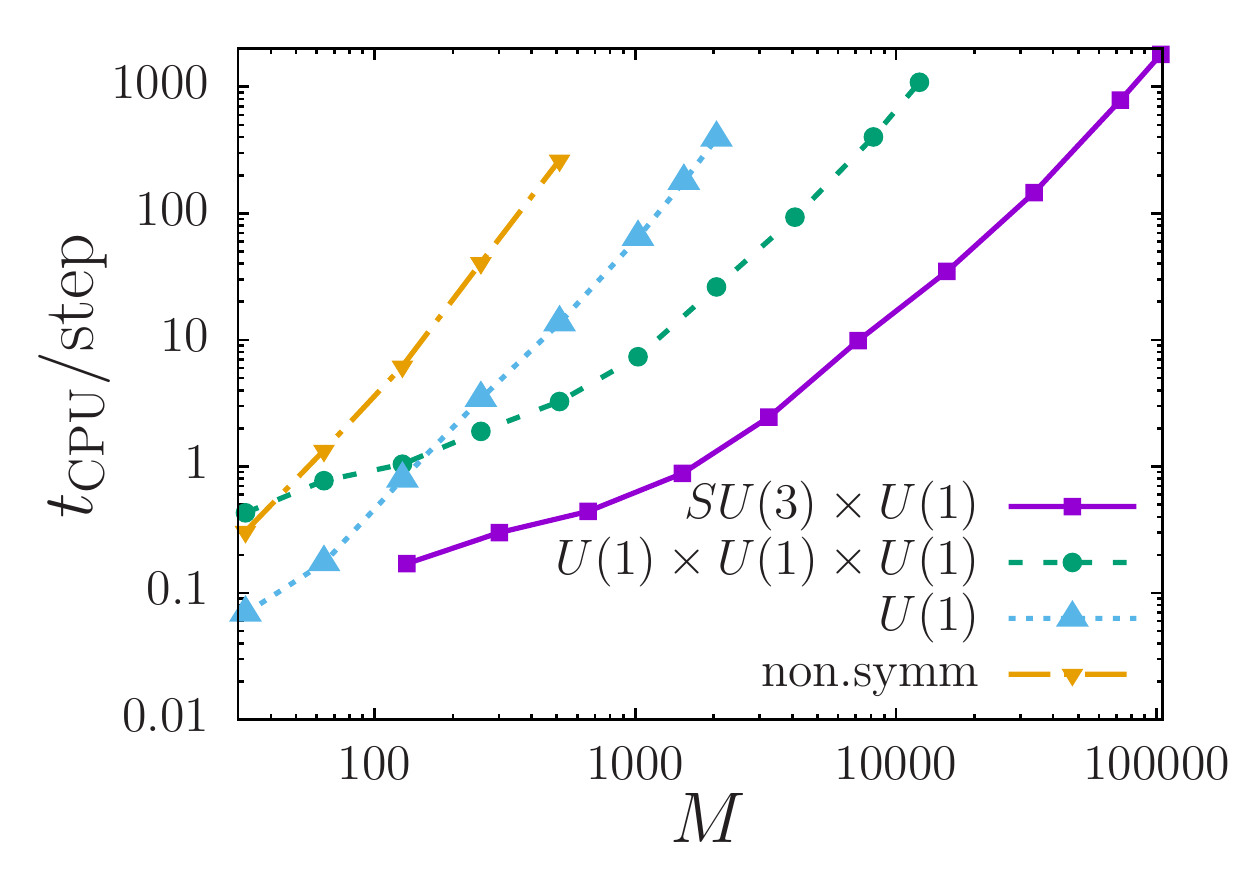}
 \caption{\label{fig:runtime}
CPU times of NA-MPS as a function of effective bond dimension, $M$, for calculations 
 exploiting various symmetries. Using non-Abelian symmetries rather than Abelian ones speeds 
  up the computations by almost two orders of magnitude.}
 \end{center}
\end{figure}

The SU(3) Hubbard model provides an ideal testbed to investigate the numerical efficiency of 
NA-TEBD.  A detailed {analysis} of the run times and the memory usage 
is presented in Figs.~\ref{fig:runtime} and \ref{fig:memory}, respectively. 

Figs.~\ref{fig:runtime}  presents the CPU time as a function of effective bond dimension, $M$, 
for various  symmetry combinations used.  Using  as many  symmetries as possible 
makes our calculations tremendously efficient. 
Using just one U(1) symmetry speeds up the calculations by a factor of $\sim20$, and we can gain 
an additional factor of  $\sim20$ in speed by exploiting  the two additional U(1) symmetries. 
However, using SU(3)$\times$U(1) symmetry rather than U(1)$\times$U(1)$\times$U(1) 
increases the speed of our calculations by an \emph{additional} factor of $\sim 100$, 
yielding an overall speed-up factor of about $\sim 100,000$. 

Similar efficiency is  reached with memory storage space. With our 20 GB memory, we can reach 
bond dimensions of about $M \sim 1000 $ without symmetries, $M\sim 10,000$ if we exploit 
the U(1)$\times$U(1)$\times$U(1)  non-Abelian symmetry, but $M\sim 100,000$  if we use our 
non-Abelian approach.  To reach these latter bond dimensions with just {Abelian symmetries}, 
one would need   a memory of around { $\sim 2\,\text{TB}$}.

\begin{figure}[t]
 \begin{center}
  \includegraphics[width=1.0\columnwidth]{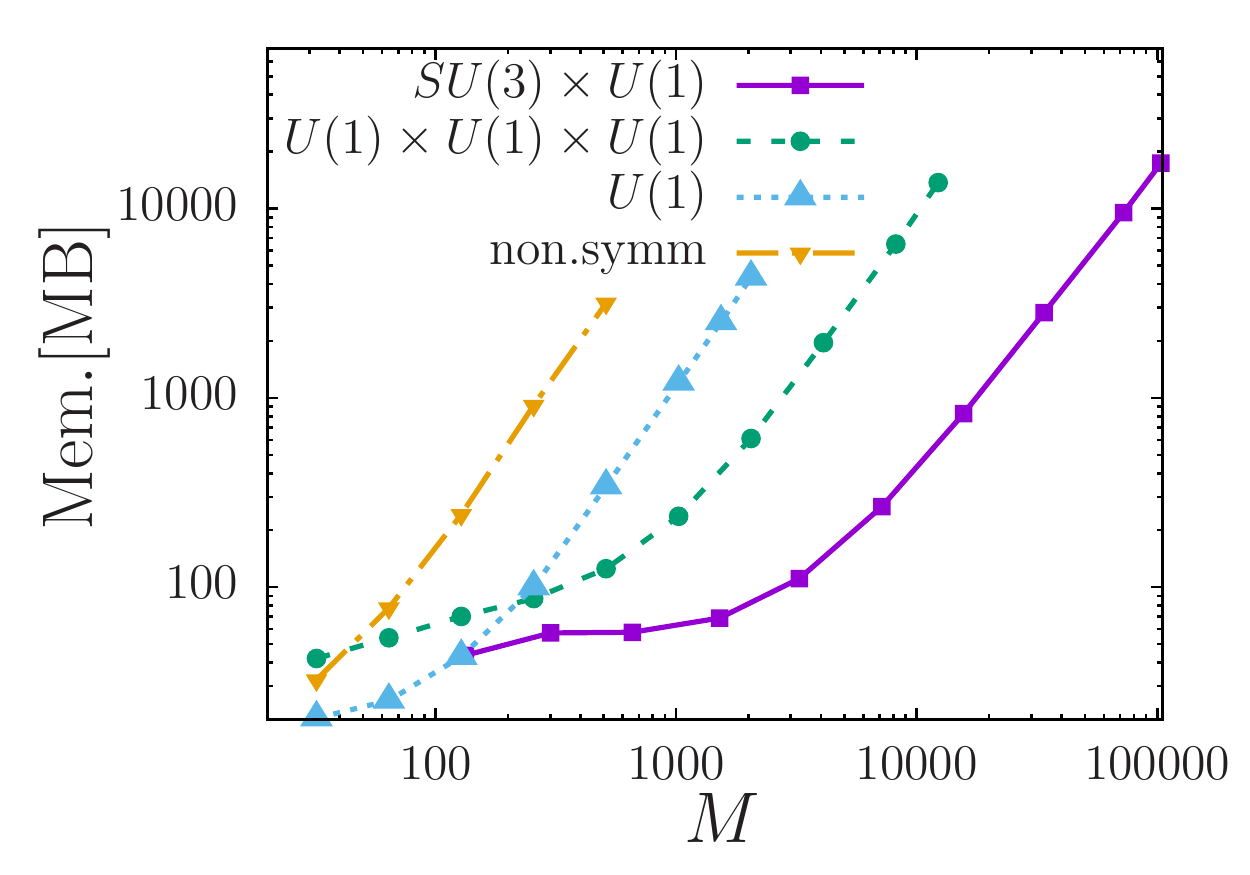}
 \caption{\label{fig:memory} Memory usage of NA-MPS as a function of effective 
 bond dimension $M$.  
  Non-Abelian symmetries reduce memory usage with respect to Abelian computations 
  by about two orders of magnitude, and allow to reach   extremely high accuravy.}
 \end{center}
\end{figure}

\section{Conclusions}
\label{sec:Conclusions}

In this work, we gave a detailed description of our   non-Abelian Matrix Product State  (NA-MPS) 
approach, which we  applied here for the SU(3) Hubbard model. {We construct the MPS state as} a two-layer 
structure, where the {'core' of the MPS, i.e.,} the first  layer is written in terms of multiplets, 
and tied through the so-called outer multiplicity labels to  a second,  Clebsch-Gordan layer. 
The latter can be consistently eliminated, thereby introducing a very efficient algorithm, where 
internal labels are suppressed. This approach leads to a 100-fold  speed-up of the 
code and a 100-fold memory reduction with respect to simple {Abelian} 
codes in case of the SU(3) Hubbard model.  We can thereby reach {extremely large} 
bond dimensions even on a small work station or even on a PC computer.
This efficiency  increase is even more dramatic for higher SU(N) symmetries, 
not studied here. This increased efficiency allows us to reach much better accuracies
{compared to codes using only Abelian symmetries}. Unfortunately, the dramatic increase 
in bond dimensions  amounts only  in a relatively small (logarithmic) increase  
in the time span of our simulations in the  particular case of the SU(3) Hubbard model.  

We then introduced and  NA-MPS based TEBD algorithm, NA-TEBD, which we used 
 to investigate charge relaxation, starting from an  initial state 
with three particles  placed at every third site of the Hubbard chain.  
For $U=0$ we derived exact  results for the single particle correlation functions, densities,  and 
the entanglement entropy, which we used to benchmark 
our direct NA-TEBD simulations.  
In the absence of interactions, we  observe algebraically decaying coherent 
charge oscillations, accompanied by a light-cone spread of correlations, and 
a linear growth of the entanglement entropy.  Remarkably large bond dimensions 
were needed to capture even the first few oscillations in this non-interacting case.

Interactions induce exponential damping  with a rate $\gamma\sim U^2$ for $U\lesssim 1$, 
indicative of a perturbative behavior for these moderate values of 
$U$.
At the same time,  the spread of correlations or the entropy growth rate remain barely affected.

The tensor structure  developed is quite rich and opens 
many possibilities to study correlations and correlated dynamics: 
we can use it to construct tree tensor network 
states (NA-TTNS), or use them  to study dissipative dynamics {through Lindbladian evolution of matrix product operators}  (NA-MPO) 
with high accuracy using non-Abelian symmetries.\cite{Moca_next} Of course, we can also extend our approach 
to perform NA-DMRG calculations or NA-DM-NRG calculations.  

\begin{acknowledgments}
This research is supported by the National Research, Development and Innovation Office - NKFIH   within the Quantum Technology National Excellence Program (Project No.
      2017-1.2.1-NKP-2017-00001), { by grant No. K120569}, and 
by the BME-Nanotechnology FIKP grant (BME FIKP-NAT). {M.A.W has also been supported by the \'UNKP-20-4-II New National Excellence Program of the National Research, Development and Innovation Office - NKFIH.}
\end{acknowledgments}

\appendix

\section{Description of non-trivial multiplets using NA-MPS}\label{appendix_sec:Nontrivial_rep}
{In the construction of NA-MPS (Section~\ref{subsection:NA-MPS}) we used the fact that for a state $ \ket {\Psi} $ belonging to the  trivial representation $\Gamma = 0$ the Schmidt-states can be sorted into multiplets. This statement is a consequence of the orthogonality theorem of group characters $ \chi_\Gamma (g) = \mathrm {tr} \brc {R _{\Gamma} (g)}$.\cite{Cornwell_vol2} The orthogonality relation reads as $ \int d \mu (g) \chi_{\Gamma_1} (g) \chi_{\Gamma_2} (g)^{*} = \delta_{\Gamma_1, \Gamma_2} $. For direct products of representations the characters are simply multiplied, therefore the orthogonality relation introduces a constraint on the Clebsch-Gordan coefficients: the trivial representation ($\Gamma = 0$) appears only in product spaces of representation -- conjugate representation pairs, with outer multiplicity one. In all other products the trivial representation is missing. As a consequence of this constraint, for a trivial state $\ket{\Psi}$ the Schmidt pairs are members of multiplets that are conjugates of each other. } 
 
{For a state $ \ket {\Psi_{\Gamma},m} $ of a non-trivial  representation $ \Gamma  \ne 0 $, it is not possible to directly write the MPS in the form of \eqref {eq:NAMPS}, since Schmidt states obtained after decomposition are not sorted into multiplets.} 
However, the problem can be circumvented by introducing an additional site that contains a multiplet for a single $ \overline {{\Gamma}} $ representation. Using this auxiliary site we define a new pure state for the whole chain as
\begin{equation}
 \ket{\tilde{\Psi}} = \sum_{m} \frac{1}{\sqrt{\dim(\Gamma)}} \ket{\Psi_{\Gamma},m} \ket{\overline{\Gamma,m}} \, , 
\end{equation}
 where $ \ket {\overline {\Gamma, m}} $ denotes the state at the auxiliary site. The $ \ket {\tilde {\Psi}} $ state defined thus belongs to the trivial representation, i.e. it can be used to build an NA-MPS. The auxiliary site is placed in the rightmost position of the chain {in our construction}.

Performing a partial trace on $ \ket {\tilde {\Psi}} $ over the auxiliary site state, we obtain 
the density matrix of the real system.
 \begin{equation}
 \hat{\rho} = \sum_{m} \frac{1}{\dim(\Gamma)} \ket{\Psi_{\Gamma},m} \bra{\Psi_{\Gamma},m} \; .
\end{equation}

\section{NA-MPS states}\label{appendix_sec:NAMPS_states}

In this section {we  present some details} on how the {reduced matrix elements, like the reduced evolver in Fig.~\ref{fig:NATEBD}} can be constructed by employing the
Clebsch-Gordan coefficients' sum rule. We start with a short description on how to 
perform the scalar product of two NA-MPSs. 
\begin{figure}[!h]
\begin{center}
 \includegraphics[width=0.9\columnwidth]{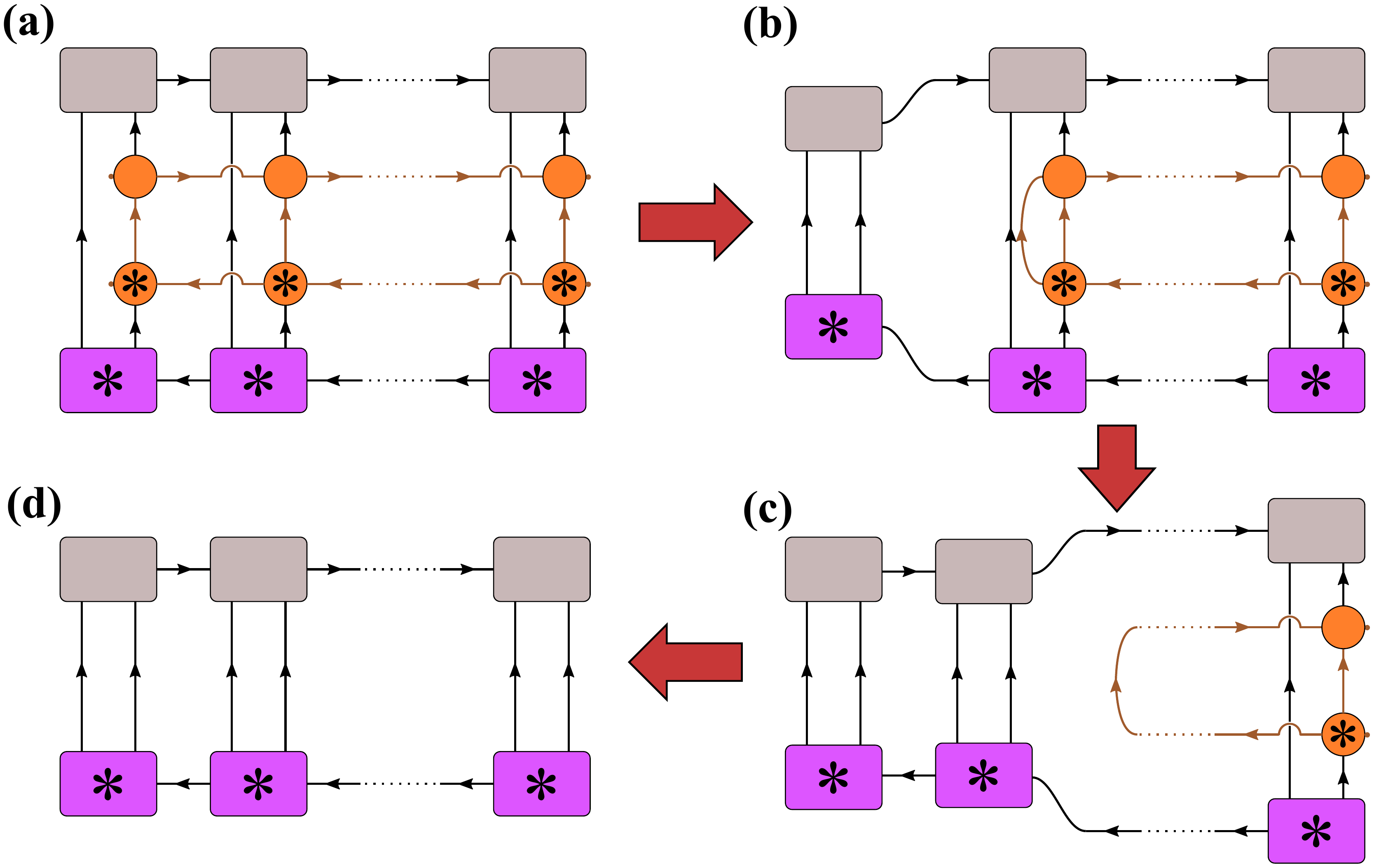}
 \caption{(a) Scalar product of two NA-MPS's. The symbol '*' attached to a tensor indicates the complex conjugation. (b) As a result of Eq.~\eqref{eq:CG_left_end}, the Clebsch-Gordan tensors (formally introduced) at the first grid location are dropped. (c-d) Visualising the contraction of the Clebsch layer by using the orthogonality equation \eqref{eq:CG_orthogonality_forward} and constructing 
 the so called  'reduced scalar product'. } \label{fig:NAMPS_dot}
\end{center}
 \end{figure}

\subsection{Scalar product}
In Fig.~\ref{fig:NAMPS_dot} (a) we introduce the graphical representation for  the scalar product of two NA-MPS's, which, by 'integrating' the Clebsches layer,  can be simplified  to 
what we call 'reduced scalar product', i.e. a scalar product 
 involving only the upper MPS layer (displayed in Fig.~\ref{fig:NAMPS_dot}.d).
 
First, we formally add a trivial site to the left belonging to the representation  $ \Gamma_0 = 0 $. The appearing 
 $ C (\Gamma_0 = 0, \Gamma^\loc_1,\Gamma_1)_{0 \, \mu_1}^{m_1 \, \alpha} $ Clebsch-Gordan 
 coefficients imply that $ \alpha $ is one-dimensional, since $ 0 \otimes \Gamma^\loc_1 $ contains only one multiplet of $ \Gamma^\loc_1 $, and furthermore
\begin{equation}\label{eq:CG_left_end}
 C(\Gamma_0=0,\Gamma^\loc_1,\Gamma_1)_{0\,\mu_1}^{m_1\,\alpha = 1} = \delta_{\Gamma^\loc_1}^{\Gamma_1} \delta_{\mu_1}^{m_1} \; .
\end{equation}
Using this equation, we graphically obtain the result presented in Fig.~\ref {fig:NAMPS_dot}.b. To move on, we use the orthogonality relation 
\begin{equation}\label{eq:CG_orthogonality_forward}
 \sum_{m,\mu} C(\Gamma,\Gamma^\loc,\Gamma')_{m\,\mu'}^{m'\,\alpha} \left(C(\Gamma,\Gamma^\loc,\tilde{\Gamma}')_{m\,\mu'}^{\tilde{m}'\,\tilde{\alpha}} \right)^{*} = \delta_{\tilde{\Gamma}'}^{\Gamma'} \delta_{\tilde{m}'}^{m'} \delta_{\tilde{\alpha}}^{\alpha}, \;
\end{equation}
to  {eliminate  locally} the Clebsch-Gordan tensors
and move site by site (a typical iteration is presented in Fig.~\ref{fig:NAMPS_dot}.c), 
to  finally  reach the right end of the chain (displayed in Fig.~\ref {fig:NAMPS_dot}.d), where the layers of $ C $ tensors have disappeared from the expression, that is, the  scalar product
of the full NA-MPSs  is given by  the 'reduced scalar product' of the upper layers.

Notice the double line structure connecting the two NA-MPS states: the first line carries the label of a local multiplet, 
while the second line carries the outer mupltiplicity labels, $\alpha$, assuring via their dependencies 
that  representation labels of corresponding bonds and local states all match in both states. 

\subsection{Matrix elements of  scalar operators}
\label{subsec:NAMPS_scalarop_matrixelement}
\begin{figure}[t]
\begin{center}
 \includegraphics[width=0.9\columnwidth]{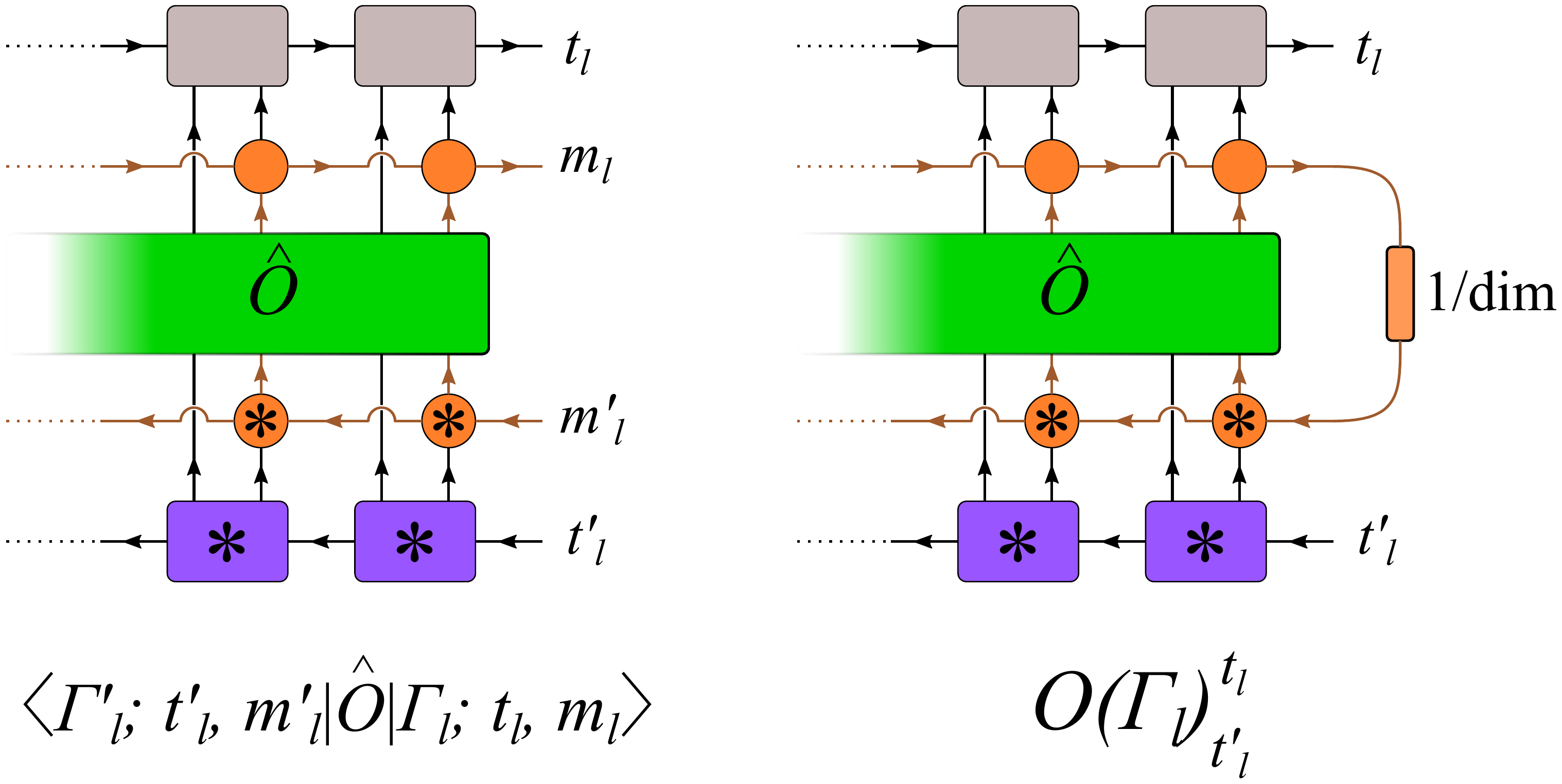}
 \caption{Left: Matrix element of a scalar operator $ \hat{O} $  between two Schmidt states.  Right: Definition of the reduced matrix element of the scalar operator as in Eq.~\eqref {eq:Scalarop_red_matelement}.}\label{fig:NAMPS_scalarop_matrixelement}
\end{center}
 \end{figure}
Let us consider an operator  $ \hat{O}$, which  only acts on a spatially localized subset of sites on the left  of 
bond $ l $. For now, we assume that  this operator is a 'scalar',  i.e., it commutes with 
all symmetry operations  $ \hat{U} (g) $. A trivial example in case of the SU(3) Hubbard model is 
the {particle} number operator, $n_l = \sum_\alpha c^\dagger_{l\alpha} c_{l\alpha}$, or any function of 
this operator. Another (not independent)  example is the {Casimir} operator ${{\cal C}_l}=\sum_i \hat \lambda_l^i \hat \lambda_l^i$. {These operators belong to the trivial representation $\Gamma=0$. Therefore acting with
them on  a Schmidt state does not change  the quantum numbers of the latter.
{In general, we can therefore write  (see also Fig.~\ref{fig:NAMPS_scalarop_matrixelement}).
\begin{equation}\label{eq:Scalarop_matelement}
\phantom{)}_l\bra{\Gamma'; t', m'} \hat{O} \ket{\Gamma; t, m}_l = \delta_{\Gamma'}^{\Gamma} \delta_{m'}^{m} O(\Gamma)_{t'}^{t} \; .
\end{equation}

 The reduced matrix elements  $ O (\Gamma_l)_{t '}^{t} $ can be easily obtained by tracing over   $ m$ and  $m' $,
\begin{equation}\label{eq:Scalarop_red_matelement}
 O(\Gamma)_{t'}^{t} =  \frac{1}{\dim(\Gamma)}  \sum_{m} \,_l\bra{\Gamma; t', m} \hat{O} \ket{\Gamma; t, m}_l \; .
\end{equation}
This relationship is shown in the right {panel} of  Fig.~\ref{fig:NAMPS_scalarop_matrixelement}.
The rectangular box  {labeled} as '1/dim' ,  represents the operation $\frac{1}{\dim(\Gamma)} \delta_{m}^{m'}$ {that again can be represented as an NA-tensor}.
Eq.~\eqref{eq:Scalarop_matelement}  is a special case of the Wigner-Eckart theorem for scalar operators. The general case is discussed in Appendix \ref{appendix_sec:Wigner_Eckart}.

\onecolumngrid

\section{The upper layer of the NA-MPS }\label{appendix_sec:NAMPS_upperlayerasMPS}
As we have {discussed}  in  Sec.~\ref{subsection:NA-MPS}, in the NA-TEBD algorithm, the upper layer of the NA-MPS can be treated in all respects as a conventional MPS state (without symmetry). We now
show in detail that the upper layer of the NA-MPS does indeed encode an MPS state, but on a very complicated basis of Hilbert space.

Consider the NA-MPS state defined in Eq. \eqref{eq:NAMPS} and represented in Fig.~\ref{fig:NAMPS}. The lower layer 
 specifies   the following product  states (complemented with the auxiliary site introduced in {Appendix~\ref{appendix_sec:Nontrivial_rep}),
\begin{eqnarray}\label{eq:abstract_basis}
  \ket{\brc{\Gamma}^{[1]}, \tau_1, \alpha_1; \; \brc{\Gamma}^{[2]}, \tau_2, \alpha_2 ; \; \dots ; \; \brc{\Gamma}^{[L]}, \tau_L, \alpha_L} &=&\sum_{\brc{\mu_l}} \sum_{\brc{m_l}} C(\brc{\Gamma}^{[1]})_{0 \; \mu_1}^{m_1 \; \alpha_1} \, C(\brc{\Gamma}^{[2]})_{m_1 \, \mu_2}^{m_2 \; \alpha_2} \dots C(\brc{\Gamma}^{[L]})_{m_{L-1} \, \mu_L}^{0 \; \alpha_L} \nonumber \\
&\times &   \ket{\Gamma^\loc_1\; \tau_1, \mu_1} \otimes \ket{\Gamma^\loc_2\; \tau_2, \mu_2} \otimes \dots \otimes \ket{\Gamma^\loc_L\; \tau_L, \mu_L} \ .
\end{eqnarray}
Here, as before, $ \brc {\Gamma}^{[l]} = (\Gamma_{l-1}, \Gamma^\loc_l, \Gamma_l) $ and $ \Gamma_0 = \Gamma_L = 0 $. These states span the {singlet sector of the} Hilbert space of the chain (extended by the auxiliary site at the right most position), and form an orthonormal basis  due to the orthogonality relation of the Clebsch-Gordan coefficients, 
Eq.~\eqref {eq:CG_orthogonality_forward}. The state notation is slightly redundant, since {representation sets} $ \brc {\Gamma}^{[l]} $ and $ \brc {\Gamma}^{[l + 1]} $ {at adjacent grid positions share the same  $ \Gamma_{l} $ representation index}. This constraint will be essential in determining the Schmidt decomposition of the {state}.
We can write an arbitrary singlet state on this   basis as
\begin{eqnarray}
 \ket{\Psi} = \sum_{\brc{\Gamma^\loc_l}} \sum_{\brc{\Gamma_l}} \sum_{\brc{\tau_l}} \sum_{\brc{\alpha_l}} & & \Psi_{\brc{\Gamma}^{[1]}, \tau_1, \alpha_1; \; \brc{\Gamma}^{[2]}, \tau_2, \alpha_2 ; \; \dots ; \; \brc{\Gamma}^{[L]}, \tau_L, \alpha_L}   \ket{\brc{\Gamma}^{[1]}, \tau_1, \alpha_1; \; \brc{\Gamma}^{[2]}, \tau_2, \alpha_2 ; \; \dots ; \; \brc{\Gamma}^{[L]}, \tau_L, \alpha_L} \nonumber \; .
\end{eqnarray}
To build the MPS, we need a Schmidt decomposition of this state, {that translates to} the SVD of the $ \Psi_{\dots} $ expansion coefficient. First we need to divide the indices into two parts. Let the cut position {be} 
between the sites $ l $ and $ l + 1 $. Then by performing the SVD of the  coefficient, we get the following expression,
\begin{multline}\label{eq:abstract_Schmidt_decomp}
 \Psi_{\left\lbrace\brc{\Gamma}^{[1]}, \tau_1, \alpha_1; \; \dots  \; ; \brc{\Gamma}^{[l]}, \tau_l, \alpha_l\right\rbrace \; \left\lbrace \brc{\Gamma}^{[l+1]}, \tau_{l+1}, \alpha_l; \; \dots \; ; \brc{\Gamma}^{[L]}, \tau_L, \alpha_L \right \rbrace} =  \\
 \phantom{asssssssssssss} \sum_{t_l} U_{\left\lbrace\brc{\Gamma}^{[1]}, \tau_1, \alpha_1; \; \dots  \; ; \brc{\Gamma}^{[l]}, \tau_l, \alpha_l\right\rbrace \; t_l} \; \Lambda^{[l]}(\Gamma_l)_{t_l} \; V_{t_l\; \left\lbrace\brc{\Gamma}^{[l+1]}, \tau_{l+1}, \alpha_{l+1}; \; \dots \; ; \brc{\Gamma}^{[L]}, \tau_L, \alpha_L \right\rbrace} \; .
\end{multline}
{The $ \brc {\Gamma}^{[l]} $ and $ \brc {\Gamma}^{[l + 1]} $ representation sets contain the common representation $ \Gamma_l $}, so by performing SVD, the  {Schmidt values} $ \Lambda^{[l]} (\Gamma_l)_{t_l} $ are also {labeled} according to $ \Gamma_l$. Eq.~\eqref{eq:abstract_Schmidt_decomp} 
is analogous to the nonsymmetric equation, {therefore} the rows of the 
$ U $ matrix and the columns of the $ V $ matrix are again orthonormal ({half-unitarity}), and {the normalization condition  $ \scalarprod {\Psi} {\Psi} = 1$ translates to} 
\begin{equation}
 \sum_{\Gamma_l} \sum_{t_l} \left| \Lambda^{[l]}(\Gamma_l)_{t_l} \right|^2 = 1.
\end{equation}
The previously introduced left-canonical $ A^{[l]} $ tensor is defined by $U$, while the 
right-canonical states can be defined in a similar way using the $V$ matrices
\begin{multline}
U_{\left\lbrace \brc{\Gamma}^{[1]}, \tau_1, \alpha_1; \; \dots  \; ; \brc{\Gamma}^{[l]}, \tau_l, \alpha_l\right\rbrace \; t_l} = \sum_{t_{l-1}} \sum_{\tau_l} \sum_{\alpha_l} U_{\left\lbrace \brc{\Gamma}^{[1]}, \tau_1, \alpha_1; \; \dots  \; ; \brc{\Gamma}^{[l-1]}, \tau_{l-1}, \alpha_{l-1} \right \rbrace \; t_{l-1}} A^{[l]}(\brc{\Gamma}^{[l]})_{t_{l-1} \, \tau_l \, \alpha_l}^{t_l} \\ 
V_{t_l\; \left\lbrace\brc{\Gamma}^{[l+1]}, \tau_{l+1}, \alpha_l; \; \dots \; ; \brc{\Gamma}^{[L]}, \tau_L, \alpha_L \right\rbrace} = \sum_{t_{l+1}} \sum_{\tau_{l+1}} \sum_{\alpha_{l+1}} B^{[l+1]}(\brc{\Gamma}^{[l+1]})_{\tau_{l+1} \, \alpha_{l+1} \, t_{l+1}}^{t_{l}} V_{t_{l+1}\; \left\lbrace\brc{\Gamma}^{[l+1]}, \tau_{l+2}, \alpha_{l+2}; \; \dots \; ; \brc{\Gamma}^{[L]}, \tau_L, \alpha_L \right\rbrace}  \; .
\end{multline}
From the half-unitarity of U and V-matrices,
 we imitedly get the half-unitarity of the $ A^{[l]} $ and $ B^{[l]} $ tensors,
 as well as the relation between them.  
\begin{eqnarray}\label{eq:left_NAMPS_orthogonality}
 \sum_{\Gamma_{l-1}} \sum_{\Gamma^\loc_l} \sum_{t_{l-1}} \sum_{\tau_l} \sum_{\alpha_l} A^{[l]}(\brc{\Gamma}^{[l]})_{t_{l-1} \, \tau_l \, \alpha_l}^{t_l} \left(A^{[l]}(\brc{\Gamma}^{[l]})_{t_{l-1} \, \tau_l \, \alpha_l}^{t_l'} \right)^{*} &=& \delta_{t_l'}^{t_l} \; \\
 \sum_{\Gamma_{l+1}} \sum_{\Gamma^\loc_l} \sum_{t_{l+1}} \sum_{\tau_{l+1}} \sum_{\alpha_{l+1}}  B^{[l+1]}(\brc{\Gamma}^{[l+1]})_{\tau_{l+1} \, \alpha_{l+1} \, t_{l+1}}^{t_{l}} \left(  B^{[l+1]}(\brc{\Gamma}^{[l+1]})_{\tau_{l+1} \, \alpha_{l+1} \, t_{l+1}}^{t_{l}'} \right)^{*} &=& \delta_{t_l'}^{t_l} \;\\
 A^{[l]}(\brc{\Gamma}^{[l]})_{t_{l-1} \, \tau_l \, \alpha_l}^{t_l} \; \Lambda^{[l]}(\Gamma_l)t_l = \Lambda^{[l-1]}(\Gamma_{l-1}) \; B^{[l]}(\brc{\Gamma}^{[l]})_{ \tau_l \, \alpha_l \, t_{l}}^{t_{l-1}} & &
\end{eqnarray}
These equations are completely analogous to the orthogonality equations Eq.~\eqref {eq:MPS_nonsym_orthogonality}  for the first, left-canonical matrix.

To conclude this appendix, we can say that the upper layer of the NA-MPS state can be understood as a conventional MPS wave function interpreted on a {basis} defined in Eq.~\eqref{eq:abstract_basis}. The resulting Schmidt weights and the properties of the left and right canonical tensors are similar to those of the regular MPSs, so 
adapting already developed algorithms to our NA-MPS  and eliminating the Clebsch-Gordan layer
does not require conceptual modifications.

\section{Details of the implementation of NA-TEBD}\label{appendix_sec:NATEBD}
In Sec.~\ref{subsec:NATEBD} we briefly introduced the basic steps  to implement the NA-TEBD algorithm. In this Appendix,
 we present the technical details of the implementation.  Let's first consider the equation defining the reduced evolver $ U_{\mathrm {red}} $  graphically defined in  Fig.~\ref {fig:NATEBD},
\begin{multline}
U_{\mathrm{red}} \left( \Gamma_{l-1}, \Gamma^{\loc}_l, \Gamma_l, {\Gamma^{\loc}_l}', \Gamma_l', \Gamma^{\loc}_{l+1}, {\Gamma^{\loc}_{l+1}}', \Gamma_{l+1} \right)_{\tau_l' \, \alpha_l' \; \tau_{l+1}', \alpha_{l+1}'}^{\tau_l \, \alpha_l \; \tau_{l+1} \, \alpha_{l+1}} = 
 \sum_{m_{l-1}} \sum_{\mu_l} \sum_{m_l} \sum_{\mu_{l+1}} \sum_{m_{l+1}} \sum_{\mu_l'} \sum_{m_{l}'} \sum_{\mu_{l+1}'} \frac{1}{\dim(\Gamma_{l+1})} \times  \\ 
 C(\Gamma_{l-1}, \Gamma^{\loc}_l, \Gamma_l)_{m_{l-1} \, \mu_{l}}^{m_l \, \alpha_l} \; C(\Gamma_l, \Gamma^\loc_{l+1}, \Gamma_{l+1})_{m_l \, \mu_{l+1}}^{m_{l+1} \, \alpha_{l+1}} \;
 U(\Gamma^\loc_l, \Gamma^\loc_{l+1}, {\Gamma^\loc_{l}}', {\Gamma^\loc_{l+1}}')_{\tau_l' \, \mu_l' \; \tau_{l+1}', \mu_{l+1}'}^{\tau_l \, \mu_l \; \tau_{l+1} \, \mu_{l+1}} \times  \\
 \left( C(\Gamma_{l-1}, {\Gamma^{\loc}_l}', \Gamma_l')_{m_{l-1} \, \mu_{l}'}^{m_l' \, \alpha_l'} \right)^{*} \left(C(\Gamma_l', {\Gamma^\loc_{l+1}}', \Gamma_{l+1})_{m_l' \, \mu_{l+1}'}^{m_{l+1} \, \alpha_{l+1}'} \right)^{*} \; .
\end{multline}
We now explicitly display   representation indices that label the blocks of each tensor. As stated  in Sec.~\ref{subsec:NATEBD}, 
the blocks of the reduced tensor $ U_{\mathrm {red}} $  have a total of eight representation indices. 

We want to formulate  TEBD  for purely left-canonical MPS's, but this requires some tricks \cite{Vidal2007}, 
since SVD always results in a left-canonical and a right-canonical tensor. The algorithm can 
be  constructed  in four steps:

\begin{enumerate}
\item Contract the tensors $ A^{[l]}$ and   $ A^{[l+1]}$ for the two neighboring sites,

\begin{equation}
W(\Gamma_{l-1},\Gamma^\loc_l,\Gamma_l,\Gamma^\loc_{l+1},\Gamma_{l+1})_{t_{l-1} \, \tau_l \, \alpha_l \, \tau_{l+1} \, \alpha_{l+1}}^{t_{l+1}} =
 \sum_{t_l} A^{[l]}(\Gamma_{l-1},\Gamma^\loc_l,\Gamma_l)_{t_{l-1} \, \tau_l \, \alpha_l}^{t_l} A^{[l+1]}(\Gamma_l,\Gamma^\loc_{l+1},\Gamma_{l+1})_{t_l \, \tau_{l+1} \, \alpha_{l+1}}^{t_{l+1}} \; .
\end{equation}

\item Construct the time evolved tensor,
\begin{multline}
\widetilde{W}(\Gamma_{l-1},{\Gamma^\loc_l}',{\Gamma_l}',{\Gamma^\loc_{l+1}}',\Gamma_{l+1})_{t_{l-1} \, \tau_l' \, \alpha_l' \, \tau_{l+1}' \, \alpha_{l+1}'}^{t_{l+1}} =  \sum_{\Gamma^\loc_l} \sum_{\Gamma^\loc_{l+1}} \sum_{\Gamma_l} \sum_{\tau_l} \sum_{\alpha_l} \sum_{\tau_{l+1}} \sum_{\alpha_{l+1}} \times \\
  U_{\mathrm{red}} \left( \Gamma_{l-1}, \Gamma^{\loc}_l, \Gamma_l, {\Gamma^{\loc}_l}', \Gamma_l', \Gamma^{\loc}_{l+1}, {\Gamma^{\loc}_{l+1}}', \Gamma_{l+1} \right)_{\tau_l' \, \alpha_l' \; \tau_{l+1}', \alpha_{l+1}'}^{\tau_l \, \alpha_l \; \tau_{l+1} \, \alpha_{l+1}} 
 W(\Gamma_{l-1},\Gamma^\loc_l,\Gamma_l,\Gamma^\loc_{l+1},\Gamma_{l+1})_{t_{l-1} \, \tau_l \, \alpha_l \, \tau_{l+1} \, \alpha_{l+1}}^{t_{l+1}} \; .
\end{multline}
\item On the right, we multiply by the appropriate Schmidt weights,
\begin{multline}
\tilde{\Theta}(\Gamma_{l-1},{\Gamma^\loc_l}',{\Gamma_l}',{\Gamma^\loc_{l+1}}',\Gamma_{l+1})_{t_{l-1} \, \tau_l' \, \alpha_l' \, \tau_{l+1}' \, \alpha_{l+1}'}^{t_{l+1}} = 
 \widetilde{W}(\Gamma_{l-1},{\Gamma^\loc_l}',{\Gamma_l}',{\Gamma^\loc_{l+1}}',\Gamma_{l+1})_{t_{l-1} \, \tau_l' \, \alpha_l' \, \tau_{l+1}' \, \alpha_{l+1}'}^{t_{l+1}} \Lambda^{[l+1]}(\Gamma_{l+1})_{t_{l+1} }\, .
\end{multline}
This step is essential for numerical stability. 
\item Execute the SVD on the $ \tilde {\Theta} $ tensor. We can do this separately for each block in $ {\Gamma_l} '$,
\begin{multline}
\tilde{\Theta}(\Gamma_{l-1},{\Gamma^\loc_l}',{\Gamma_l}',{\Gamma^\loc_{l+1}}',\Gamma_{l+1})_{t_{l-1} \, \tau_l' \, \alpha_l' \, \tau_{l+1}' \, \alpha_{l+1}'}^{t_{l+1}} \Rightarrow 
 \sum_{t_l'}  \tilde{A}^{[l]}(\Gamma_{l-1},{\Gamma^\loc_l}',{\Gamma_l}')_{t_{l-1} \, \tau_l' \, \alpha_l'}^{t_l'} \times \\
 \phantom{abcdefghijkl} \tilde{\Lambda}^{[l]}({\Gamma_l}')_{t_l'} \tilde{B}^{[l+1]}({\Gamma_l}', {\Gamma^\loc_{l+1}}',\Gamma_{l+1})_{t_l' \, \tau_{l+1}' \, \alpha_{l+1}'}^{t_{l+1}} \; .
\end{multline}
\item The new $ \tilde {A}^{[l + 1]} $ tensor for the right site is obtained from $ \widetilde {W} $ by utilizing the orthogonality equation, 
 Eq.~\eqref{eq:left_NAMPS_orthogonality}, 
\begin{multline}
 \tilde{A}^{[l+1]}({\Gamma_l}', {\Gamma^\loc_{l+1}}',\Gamma_{l+1})_{t_l' \, \tau_{l+1}' \, \alpha_{l+1}'}^{t_{l+1}} = \sum_{\Gamma_{l-1}} \sum_{{\Gamma^\loc_l}'} \sum_{t_{l-1}} \sum_{\tau_l'} \sum_{\alpha_l'} 
   \widetilde{W}(\Gamma_{l-1},{\Gamma^\loc_l}',{\Gamma_l}',{\Gamma^\loc_{l+1}}',\Gamma_{l+1})_{t_{l-1} \, \tau_l' \, \alpha_l' \, \tau_{l+1}' \, \alpha_{l+1}'}^{t_{l+1}} \times \\ 
 \left( \tilde{A}^{[l]}(\Gamma_{l-1},{\Gamma^\loc_l}',{\Gamma_l}')_{t_{l-1} \, \tau_l' \, \alpha_l'}^{t_l'} \right)^{*}
\end{multline}
\end{enumerate}

\twocolumngrid
Apparently, the algorithm can be implemented using purely left-canonical $ A^{[l]} $ tensors, but we also need to store and update the $ \Lambda^{[l]} (\Gamma_l)_{t_l} $ Schmidt weights, provided by the SVD step 4.

\section{Handling non-scalar operators}\label{appendix_sec:Wigner_Eckart}
In the NA-TEBD algorithm, we have seen that the reduced shape of a scalar operator belonging to two adjacent lattice sites can be easily determined by contractions with Clebsch-Gordan tensors. However, this method is difficult to generalize for handling distant interactions, or for calculating distant correlations, since the reduced coupling contains all the lattice locations between interacting lattices at once, meaning that we would store a huge multi-lattice operator, which quickly leads to depletion of computing and storage capacities. This problem can be circumvented by generalizing Eq.~\eqref {eq:Scalarop_matelement}, which is possible by the Wigner-Eckart theorem. For this we need the notion of irreducible tensor operators (henceforth simply tensor operators). These are operator multiples 
of $ \hat {O} ({\Gamma_{\mathrm {op}}})^{M} $, with  ($ M \in \lbrace 1 \dots \dim {\Gamma_{\mathrm {op} }}) \rbrace $), which are transformed by $ \hat {\mathcal {U}} (g) $ symmetry transforms as represented by $ \Gamma_{\mathrm {op}} $, as follows,
\begin{equation}\label{eq:tensor_operator}
 \hat{\mathcal{U}}(g) \; \hat{O}({\Gamma^{\mathrm{op}}})^{M}  \; \hat{\mathcal{U}}(g)^\dag = \sum_{M'} \left[R_{\Gamma^{\mathrm{op}}} \right]_{M'}^{M} \hat{O}(\Gamma^{\mathrm{op}})^{M'} \; .
\end{equation}
As an example, consider the standard spin operator, which is the combination of the three spin components ($ \hat {S}^x $, $ \hat {S}^y $, $ \hat {S}^z $). The spin operators form a three-dimensional ($ S_\mathrm {op} = 1 $) {multiplet}, whose elements
\begin{equation}
 \hat{S}^{M} = \left(-\hat{S}^{+}/{\sqrt{2}} \, , \; \hat{S}^z \, , \; \hat{S}^{-}/{\sqrt{2}} \right) \; ,
\end{equation}
 provided that the state space is expressed in the basis of the eigenvalues of the spin component $ \hat {S}^z $. Here $ \hat {S}^{\pm} = \hat{S}^{x} \pm i \hat {S}^y $ are the usual spin-shift operators.
 
The Wigner-Eckart theorem follows from the observation that if we act on the states of a multiplet of a representation $ \Gamma $  with elements of the operator multiplet of $ \Gamma_{\mathrm {op}} $, the result will transform under the product representation $ \Gamma \otimes \Gamma_\mathrm {op} $. This product can be grouped again into multiplets using the Clebsch-Gordan coefficients. Therefore, for the matrix elements of the tensor operators we obtain the following equation,
\begin{widetext}
\begin{equation}\label{eq:Wigner_Eckart}
 \bra{\Gamma'; t',m'} \hat{O}(\Gamma^\mathrm{op})^{M} \ket{\Gamma; t,m} = \sum_{\alpha} \mathbb{O}(\brc{\Gamma})_{t'}^{t \; \alpha} \left( C(\brc{\Gamma})_{m \, M}^{m' \, \alpha} \right)^{*} \; ,
\end{equation}
\end{widetext}
where $ \brc {\Gamma} = (\Gamma, \Gamma^\mathrm {op}, \Gamma') $ represents the {representation} indices that appear. Comparing this with Eq.~\eqref {eq:Scalarop_matelement} we notice that $ \mathbb {O} {\brc {\Gamma}}_{t'}^{t \; \alpha} $ is a reduced matrix element, but it contains three different {representation} indices for general tensor operators and {an $ \alpha $ outer-multiplicity index that is contracted with the Clebsch-Gordan tensor}.

The reduced matrix element can be obtained from Eq.~\eqref{eq:Wigner_Eckart}  using the orthogonality relation Eq.~\eqref{eq:CG_orthogonality_forward} \,
\begin{widetext}
\begin{equation}\label{eq:Wigner_Eckart_redmatelement}
 \mathbb{O}(\brc{\Gamma})_{t'}^{t \; \alpha} = \sum_{m,m',M} \frac{1}{\dim(\Gamma')} \; C(\brc{\Gamma})_{m \, M}^{m' \, \alpha} \; \bra{\Gamma'; t',m'} \hat{O}(\Gamma_\mathrm{op})^{M} \ket{\Gamma; t,m} \; .
\end{equation}

\end{widetext}
\bibliography{references}

\end{document}